\documentclass[fleqn,usenatbib]{mnras}

\usepackage{newtxtext,newtxmath}

\usepackage[T1]{fontenc}

\DeclareRobustCommand{\VAN}[3]{#2}
\let\VANthebibliography\thebibliography
\def\thebibliography{\DeclareRobustCommand{\VAN}[3]{##3}\VANthebibliography}

\usepackage{graphicx}	
\usepackage{amsmath}	

\usepackage{xcolor}

\newcommand\cf[1]{\textcolor{blue}{#1}}

\def\las{\mathrel{\hbox{\rlap{\hbox{\lower3pt\hbox{$\sim$}}}\hbox{\raise2pt\hbox{$<$}}}}}

\title[A Planet as the Possible Cause of the HD~181327 Debris Disk Asymmetry]{A Planet as the Possible Cause of the HD~181327 Debris Disk Asymmetry}

\author[Fox, Wiegert]{Chris Fox$^{1,2}$\thanks{Contact e-mail: \href{mailto:cfox53@uwo.ca}{cfox53@uwo.ca}}, Paul Wiegert$^{1,2}$
\\
$^{1}$Department of Physics \& Astronomy, The University of Western Ontario, London, Ontario, Canada\\
$^{2}$Institute for Earth and Space Exploration (IESX), The University of Western Ontario, London, Ontario, Canada}

\date{Accepted to the Monthly Notices of the Royal Astronomical Society, August 2026}

\pubyear{2026}

\begin{document}
\label{firstpage}
\pagerange{\pageref{firstpage}--\pageref{lastpage}}
\maketitle

\begin{abstract}
The debris disk around HD~181327 shows a significant asymmetry in its surface brightness profile when viewed in visible light.  Observations from the Hubble Space Telescope STIS instrument show an arc of approximately 90$^\circ$ of higher optical depth at a distance of 84~au from the star.  Here we explore whether a planet could be the driver of this asymmetry.  Through N-body simulations, we find that a 2-4 Jupiter-mass planet on a circular orbit at 70~au can produce and maintain a similar feature if the collisional lifetime of dust in the disk is between 4~and~8~kyr, so long as a modest fraction of the grains are ejected with high ($\lesssim$1~km/s) speeds.  We also find that the surface brightness asymmetry is much less pronounced at larger particle sizes, which may account for the fact that observations of HD~181327 at longer wavelengths have not reported such an arc. 
\end{abstract}



\section{Introduction}
HD~181327 (also known as HIP 95270) is an 18.5 Myr old main sequence F5/6 star in the Beta Pictoris Moving Group \citep{milli2024, miretroig-betapic2020}.  Its debris disk was first directly observed by \cite{schneider2006} in the near-infrared, and the disk has been the target of many further observations since.  \citet{lebreton2012}'s examination of it in the far-infrared revealed the presence of icy plantessimals.  \citet{marino2016} reported the presence of exocometary gas in the disk.  \citet{milli2024} examined polarization properties of the disk and found evidence for sub-micron particles.  Most recently \cite{Xie2025}, using data from the James Webb Space Telescope, found evidence for crystalline water ice and continual collisions within its debris disk.

Most relevant to our work here, \citet{stark2014} observed the disk with the Space Telescope Imaging Spectrograph (STIS) on board the Hubble Space Telescope.  They found a debris disk with optical depth peaking at 84.2~au from the central star.  When deprojected, the disk showed a significant azimuthal asymmetry manifesting as a ~90$^\circ$ arc of increased optical depth.  In that work, the asymmetry was interpreted as a region of higher particle density, with the cause hypothesized as warping due to interaction with the interstellar medium or the result of a catastrophic disruption of a 1\% Pluto-mass object within an exo-Kuiper belt. In a related work, \citet{schneider2014} suggested the asymmetry may indicate the presence of unseen planetary companions.

It has long been recognized that debris disks may host unseen planets  \citep{wyatt1999} and that planets may produce observable disk structures while the planets themselves may remain unobserved \citep{Mouillet_1997, Augereau_2001, Kraus_2013, tabwie16, tabwie17, tabwie18}.  This is also seen in our own solar system, where planets form structures by trapping grains in mean motion resonances \citep{dermott1994,liou1999}.

Could the visible asymmetry of the debris disk of HD~181327 be created by the gravitational perturbations from a planet?  Here we explore this premise through the use of N-body simulations of a hypothetical planet on the inner edge of the HD~181327 disk and constrain the orbital parameters of such a planet.

\section{Methodology}
The primary goal of this paper is to determine whether the presence of a planet near the inner edge of the HD~181327 disk provides a plausible explanation for the structures observed by \citet{stark2014}.  In that paper, raw STIS observations were deprojected to produce a face-on map of the normalized optical density, resulting in Figure~6h from that paper. That map of optical density serves as the basis for our work, where we use a Bayesian search coupled iteratively with a numerical integrator to find the planetary parameters that produce a disk that most closely matches the observations.  Details of this process are discussed in the following sub-sections.

\subsection{Update of Distance Values}
\citet{stark2014} reported the peak of the HD~181327 ring at a distance of 90.5~au from the central star.  That work (and several others) use a distance for the system of 51.8~pc.  However, updated values from Gaia \cf{DR2} put the stellar distance of HD~181327 at \cf{48.21$\pm$0.13~pc \citep{GaiaCollabDR2}}, and our work will be performed using this value.  Rescaling Stark's distance with the newer distance estimate results in a ring radius of 84.2~au.  We scale all of Stark's (and other sources) distance values throughout this paper accordingly.

\begin{figure}
 \includegraphics[scale=0.35]{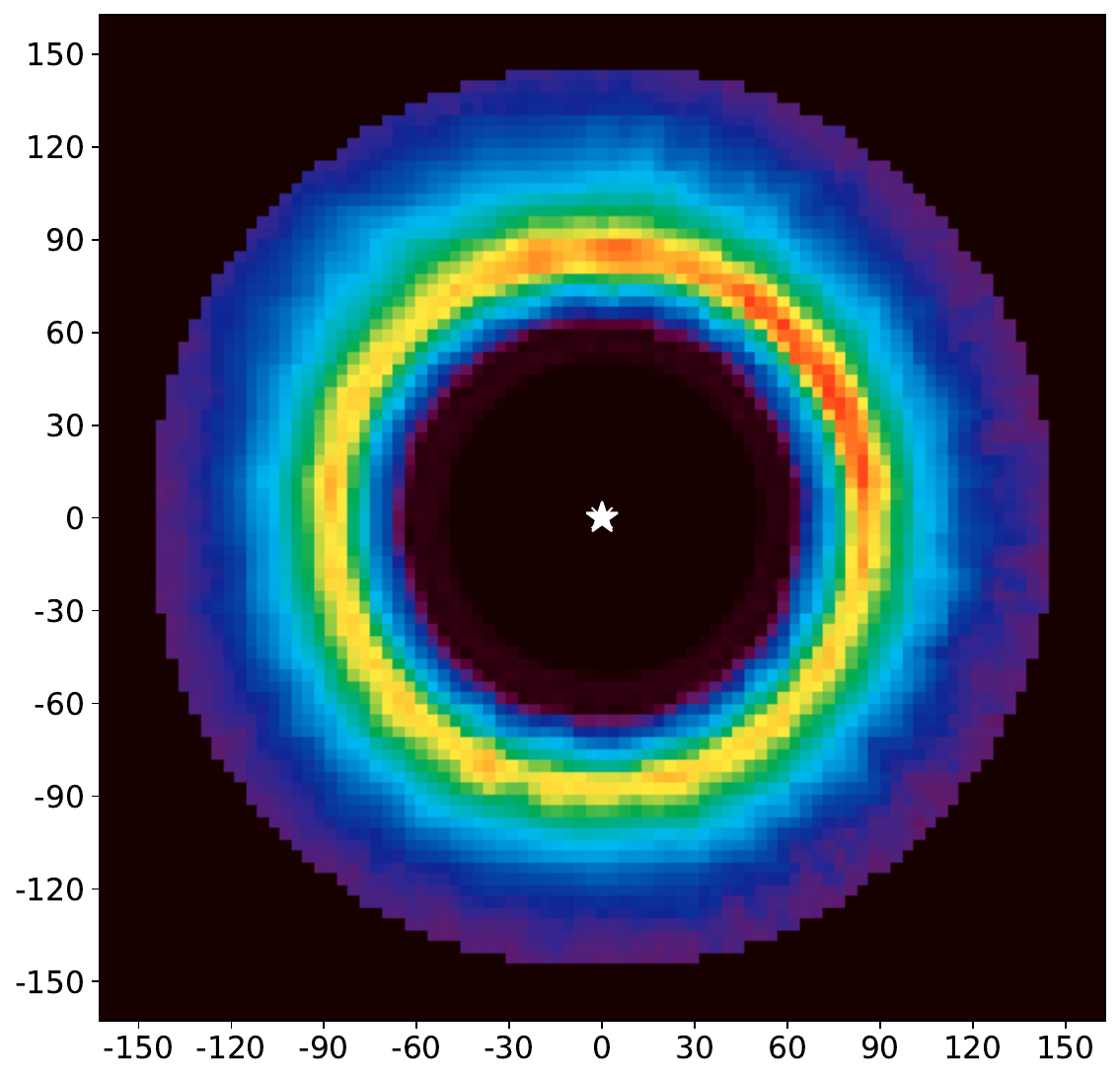}
 \includegraphics[scale=0.25, trim=100 -593 0 0]{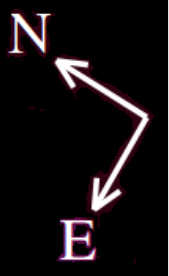}
 \includegraphics[scale=0.51, trim=0 -20 0 0]{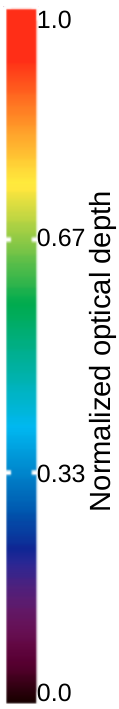}
 \caption{\label{fig:targetmap89shift} Our revised Target Map of HD~181327, used for all comparisons in our simulations.  This is based on the original from \citet{stark2014}, but with data outside of 150~au removed.  This is the map our simulations aim to replicate via the influence of a shepherding planet.  The arc is near-centred on the western side. }
\end{figure}

\subsection{Creating the Target Map}
Our first step was to adapt Figure~6h of \citet{stark2014} to our purposes.  This is the deprojected, normalized, optical depth map of the STIS image.  We extracted a subset of Stark's original map, sized to cover the majority of the disk but excluding the faint edges well beyond the visible asymmetries.  This corresponds to an outer radius of 150~au.  Our resulting map extent is 89$\times$89 pixels, with each pixel being equivalent to 3.646~au across.  We chose to use the same colour scale for optical depth values as Stark's.  For each pixel, we found the colour hex code of Stark's image, matched that to the hex code value of the colour scale, and recorded the corresponding optical depth value. 

Our revised map, adapted from the original Figure~6h from Stark, is seen in Figure~\ref{fig:targetmap89shift} where we have adopted the same colour scale as Stark et al.  Henceforth, we refer to this as "Target Map" or "Target Disk", which we aim to replicate via our simulations.

\subsection{Simulation Initial Conditions}
\subsubsection{Dust grain sizes}
\label{sec:dustgrainsizes}
Our simulations examine particles of different sizes, their dynamics being characterized by $\beta$, the ratio of the force from stellar radiation to the star's gravitational force.  $\beta$ may be expressed as \citep{wyattwhipple1950}:
\begin{eqnarray}
  \label{eq:beta}
    \beta = \frac{F_{rad}}{F_{grav}} = \frac{3 {L_{\star}} Q_{pr}}{16 \pi G {M_{\star}} c \rho s} 
\end{eqnarray}
where $c$ is the speed of light, $G$ is the gravitational constant, $L_{\star}$ is the luminosity of the star, and $Q_{pr}$ is the radiation pressure coefficient.  For HD~181327, Equation~\ref{eq:beta} becomes 
\begin{eqnarray}
  \label{eq:betashort}
    \beta = 1.396 \left(\frac{s}{\micron} \right)^{-1} \left( \frac{\rho}{\rm{g/cm^{3}}} \right)^{-1} 
\end{eqnarray}
where $\rho$ is the grain density, $s$ is the grain radius, and $Q_{pr}$ is assumed to be unity.

\citet{lebreton2012} attempted to constrain the composition and size distribution of the HD~181327 disk.  They found the best fit minimum grain size of 0.9$\micron$. \citet{milli2024} also found evidence for sub-micron grains.  The best model from \citet{lebreton2012} consisted of a disk composed of porous amorphous silicates and carbonaceous material and ice, suggesting low density (<1$\rm~g/cm^{3}$) particles.  However, more recently \citet{milli2024} found that their observations required a more refracting component such as iron-bearing material, which would have a significantly higher density.  

For this paper, we are only concerned with the dynamical behaviour of the particles, not their exact composition or sizes.  For convenience we assume a mean density of 2$\rm~g/cm^{3}$, similar to our own solar system.  With this density, our range of examined $\beta$ values from 0.05 to 0.8 would correspond to a physical radius range of 14 to 0.8 $\micron$ respectively, which are effective light scatterers at STIS wavelengths: a central bandpass of 0.585~$\micron$ with a FWHM of 0.441~$\micron$ \citep{hstSTIS}.  Even if our assumed typical grain density is 4$\times$ too low, the particles for this range of $\beta$ would still be effective scatterers in optical wavelengths.

\subsubsection{Initial Particle Distribution and Parameters} \label{sec:partdist}
Here we first describe a simplified particle size and speed distribution adopted as a result of positive results in initial experiments, and then assess our model's sensitivity to that distribution in Section~\ref{sec:refinedmodel}. 

The disk of HD~181327 consists of a central "parent ring" and an extended outer cloud.  The parent ring is a relatively narrow annulus of ongoing collisions \citep{Xie2025}.  For our simulations, this region served as the source of all particles.  Each of these particles was assigned a random $\beta$ value between 0.05 and 0.8.  We explored various subsets of particles within this range.  The particles are initially uniformly distributed within an annulus with a speed and direction appropriate to a circular orbit.  To represent the grain's creation due to a collision, a random kick velocity (using sphere-point picking method) is then applied.  Particles that move out of the central ring at the farther limits of their orbits, bound and unbound, populate the outer cloud.

During our initial simulations and experiments, we found the best match to the Target Disk when a substantial component of the dust had $\beta$$\approx$0.5. Particles with $\beta$>0.5 are often assumed to be quickly lost from the system ("beyond the blow-out threshold") because a particle with $\beta$>0.5 released at zero speed from a parent body on a circular orbit will immediately find itself on an unbound orbit and leave the system. However, particles with $\beta$>0.5 can remain bound to the star if they are given sufficient ejection speed and direction from the planetesimal.

To simulate this, the particles were initially given an eccentricity of 1.0, matching what a $\beta$=0.5 particle would have if released from the planetesimal with no relative velocity.  These particles were then given a random "kick": a burst of velocity in a random direction (including the z-axis) to represent their explosive non-isotropic ejection as the result of a collision.  Most particles are ejected with low speeds, and thus remain on hyperbolic orbits.  These particles are lost in just a few hundred years.  The few particles with sufficient ejection speed and direction to put them into bound orbits are retained, and thus contribute more to the visible disk than the ejected particles in the long-term.

For the case of HD~181327, to put a $\beta$$\approx$0.5 particle into a circular orbit the ejection speed required is 1.1~km/s relative to the parent body.  The orbital speed at 84.2 au is $v_{orb}$=3.8~km/s (for a circular orbit).  Since particle ejection speeds reach maximum values comparable to the impact speed \citep{houshols2011, jutzi2019}, and impact speeds go like $\langle e \rangle v_{orb}$ where $\langle e \rangle$ is the average eccentricity \citep{lohkrirod08}, we can expect ejection speeds to reach the needed values if <$e$>$\sim$1.1/3.8 = 0.29.  We will find that such eccentricities are common within our simulation.  Since high-speed ejecta is only a small fraction of the total released by an impact, production of a sufficient amount of high-speed material remains a potential difficulty of the planetary model adopted here.  The presence of the planet itself provides a natural source (via gravitational scattering) of higher eccentricity particles, and higher collision speeds, further creating more high-eccentricity particles.  Nonetheless, we proceeded on the premise that our collisions of $\sim$1~km/s could similarly produce ejecta of $\sim$1~km/s, and reexamined this issue later. (See Sections~\ref{sec:combanalysis} and \ref{sec:modellimits}.) 

The random kick of this magnitude is found to leave a collection of particles with many eccentricity values, averaging to $\sim$0.3.  The relative speed of the particles with respect to planetesimals is of order $e \times v_{orb}\sim$1.4~km/s (and given the large range of eccentricities seen later, even higher relative speeds are possible), so our model is self-consistent with respect to the collision/ejection speeds expected.  Thus in our model while the disk appears circular, it primarily consists of particles with eccentricities of 0.2 to 0.8 (see results in Section~\ref{sec:resultdist}) and it is this configuration that produces a continual creation of particles with sufficient collisional and ejection speeds.

\subsubsection{Particle Count and Stacking}
Our simulated disk begins with 4800 particles, and another 4800 are added at each timestep, representing the creation of new particles.  Snapshots of the disk are taken at 100 evenly spaced time intervals throughout the simulation, which are then stacked together.  The total length of the simulation thus corresponded to the collisional lifetime of the dust and includes particles of a range of ages; for simplicity we assumed that all the dust survives for the entire simulation and is destroyed at the end. The stacked snapshots thus represent the quasi-equilibrium state resulting from the continual creation, evolution and destruction of particles through their lifetime.  The number of particles used in the final comparison could potentially be as high as 4800$\times$101=492880.  However, due to radiation pressure and the applied ejection kick, more than half of newly created particles are lost in just a few hundred years as they are immediately put on unbound orbits.  During the entire evolution of the disk about 80\% of them are lost (see Sections \ref{sec:partdist} and \ref{sec:lifetimes}).  After stacking, the total particle count is typically around 200,000.

\subsection{Running the Simulations: Numerical Integrator}
\label{sec:sisyphus}
The numerical integrator we use for modeling the disk is based on the Wisdom-Holman fast symplectic integration scheme \citep{wishol91}, with a time step of 500 days.  This time step is approximately 5\% of the planet's period when its semimajor axis is 10~au (within the lower limit of the prior, Section~\ref{sec:bayesian}).  This allows for adequate sampling of all particles throughout their orbit regardless of the planetary semimajor axis chosen. Only planets interior to the disk are considered.  Radiation pressure and Poynting-Robertson drag are included for particles with $\beta$>0. The duration of our simulations depended upon the expected collisional lifetime of the particles, described in the next section.

\subsection{Simulation Duration / Particle Lifetimes} 
\label{sec:lifetimes}The simulated disk is affected by the adopted collisional timescale of the particles, as the particles have only this amount of time to respond to the presence of the planet and develop any resulting structures.  

We can estimate the collisional lifetime of the HD~181327 disk from the particle density as determined by the amount of light received at Earth. The surface brightness of a pixel in terms of power received by Hubble's STIS instrument is:
\begin{eqnarray}
  \label{eq:hubPower}
    P_{H} = A_{H} \frac{L_{p}}{4 \pi D^{2}} 
\end{eqnarray}
where $A_{H}$ is the collecting area of Hubble's primary mirror (4 m$^{2}$), $L_{p}$ is the power reflected by the dust in that pixel, and $D$ is the distance to the star (48.2 pc).  The raw STIS data from \citet{stark2014} (Figure~1 of that paper) has a pixel width of $W=2.446$~au (this is adjusted from the original Stark value, using the updated Gaia distance), a peak value of 6 counts, and a CCD Gain of 4 \citep{schneider2014}.  Assuming an average photon wavelength of 0.585~$\micron$, the central bandpass of STIS \citep{hstSTIS}, we estimate the peak power received at Hubble from one pixel as $8.15\times10^{-18}$W.

The power reflected by one dust grain is:
\begin{eqnarray}
  \label{eq:dustLum}
    L_{grain} =  \alpha \pi s^2 \frac{ L_{*}}{4 \pi r^{2}}
\end{eqnarray}
where $L_{*}$ is the luminosity of the star, $\alpha$ is the albedo of the dust particle, $r$ is the distance of the particle from the star and $s$ is the radius of the dust particle. \citet{lebreton2012} puts the observed albedo at 0.13 which we will adopt here though this is at 1.1$\micron$, slightly above the upper wavelength limit of STIS.  

Combining Equations \ref{eq:hubPower} and \ref{eq:dustLum} we can derive the number $N$ of particles inside of one image pixel as:
\begin{eqnarray}
  \label{eq:partCount}
    N = \frac {16 \pi D^{2} r^{2} P_{H} } { \alpha L_{*} A_{H} s^{2}}
\end{eqnarray}

The surface density is the number of particles in a square pixel (width $W$=2.446~au) divided by its area.  The vertical optical depth $\tau$ (surface density $\times$ particle cross-sectional area) can be expressed as:
\begin{eqnarray}
  \label{eq:optdepth}
    \tau &=& \frac {16 \pi^{2} D^{2} r^{2} P_{H} } { \alpha L_{*} A_{H} W^{2}} = 0.005 \left( \frac{\alpha}{0.13} \right)^{-1} \left( \frac{D}{48.2~\rm{pc}} \right)^2
\end{eqnarray}
This gives us an estimated peak optical depth of $\tau=0.005$, which confirms that we are dealing with an optically thin disk.  

We now seek to estimate the collisional lifetime of a particle.  From Equation~\ref{eq:optdepth}, we estimate a particle volume density by assuming a disk scale height of 6~au  \citep{schneider2006, stark2014, milli2024}, and then estimate a mean free path of 867~au.  Combining typical orbital distance ($\approx$120~au for a $e$=0.3 particle with periastron of 84.2~au), inclination (6~au~/~84.2~au $\approx$ 4$^\circ$) and eccentricity ($\sim$0.3), the typical relative speed between crossing eccentric particles is $\sim$0.9 km/s (and can be much greater for higher eccentricities).  These particles will sweep out the mean free path in $\sim$6.1~kyr.  This represents the typical lifespan of a particle ejected from the parent disk. 

We note that the literature contains various formulae for collisional lifetimes of the smallest particles, and estimates for particle lifetimes vary significantly.  Formulae based on the angular velocity and optical depth, $t_{coll}=(\Omega\tau)^{-1}$, are frequently used to estimate the lifetime of the smallest particles \citep{wyatt1999, ThebaultAugereau2007, lebreton2012} and produce collision times for HD~181327 from as low as 7600 years to 20~kyr.  However, numerical model-derived lifetimes indicate particles smaller than 0.1~mm in a typical disk can have longer lifetimes by multiple orders of magnitude \citep{ThebaultAugereau2007}, potentially up to $\sim$1~Myr.  Most recently, \citet{Xie2025} predicted a lifetime of 3300 years, and acknowledged the difficulties with making such estimates.

Hence, given the complexities in computing particle lifetimes and our own simplifications, our simulations were performed over a range of particle lifespans: from 3~kyr to 20~kyr.

\subsection{Conversion of Simulated Particle Surface Density to Optical Depth Map}
\label{sec:convODM}
Our Target Map is an 89$\times$89 grid of optical depth values, while our simulation is comprised of individual particles.  This disk is normalized to the Target Disk in terms of the total optical depth (the sum of values of all pixels).  The disk of HD~181327 has been determined to be optically thin \citep{schneider2006, lebreton2012}, so the surface density of our simulated particle disk is simply linearly related to the optical depth of the disk. We scale the optical depth of our simulated disks so that the total optical depth of all the pixels in the simulated disk is equal to the sum of the optical depths of all the pixels in the Target Disk.

To be sure that each simulated disk is fairly compared, each simulation is rescaled according to the number of particles remaining at the end of the simulation, not just the number initialized at the beginning. For example, if in one simulation half the particles are lost by the end of the simulation, then each remaining particle has to account for twice the optical depth.

\subsection{Comparing Simulated Disks to the Target Disk: a Bayesian Search}
\label{sec:bayesian}
To find the planetary parameters that best reproduce the Target Disk, we used MultiNest, a Bayesian Inference Tool \citep{feroz2009}, coupled with our numerical integrator (Section~\ref{sec:sisyphus}).  Our MultiNest search included five parameters: the semimajor axis of the planet (assumed to be on a circular orbit), the mass of the planet, the center of the parent disk annulus, the width of the annulus (full width), and the random "kick" velocity (see Section~\ref{sec:partdist}).  We use uniform priors for all parameters.  For the mass, the prior extends from 0.1 to 13.1~$\rm M_{J}$ masses, with the upper limit being the approximate lower limit for brown dwarfs.  For the semimajor axis, the prior is 30 to 80~au; the inner chosen so it is inside the limit set by \citet{rodigas2014}, and the outer limit chosen so that it overlaps the inner edge of the Target Disk.  The priors of the parent disk center was 70 to 100 au to account for the possibility that the observed disk may evolve from a different initial location.  The parent disk width prior ranged from 5 to 35 au, designed to more than cover the observed disk and account for evolution.  Finally, the random ejection ("kick") prior had limits of 800 to 1400 m/s.  This range is based on the required speed to put a $\beta$$\approx$0.5 particle on a 0 eccentricity orbit, 1100 m/s. The priors are summarized in Table~\ref{tab:priors}. 

\begin{table}
	\centering
	\caption{Planet and Disk Priors}
	\label{tab:priors}
	\begin{tabular}{rcc} 
	      & Limits & Units \\
		\hline
		Planet mass & [ 0.1 , 13.1 ] & $\rm M_{J}$\\
        Planet semimajor axis & [ 30.0, 80.0 ] & au\\
        Parent disk center & [ 70.0, 100.0 ] & au\\
        Parent disk width (full) & [ 5.0, 40.0 ] & au\\
        Particle ejection speed & [ 800, 1400 ] & m/s\\
	\end{tabular}
\end{table}

MultiNest chooses the parameters from the prior, which are then fed into the numerical integrator.  The simulation is run and the particles at different timesteps are collected (see below).  The combined disk particles are converted to a grid of 89$\times$89 optical densities with the total optical depth scaled to the Target Disk total (see Section~\ref{sec:convODM}).  MultiNest then compares this map to the Target Disk values.  For quality of comparison, we used the log-likelihood, based on the usual $\chi^{2}$ metric.

\begin{eqnarray}
    \label{eq:logL}
    \ln{L} = \sum_{i=0}^{N} \Bigg( \Big(\ln{\frac{1}{\sigma_i\sqrt{2\pi}}\Big)} -\frac{(x_i-\mu_i)^2}{2\sigma_i^2} \Bigg)
\end{eqnarray}
In Equation~\ref{eq:logL} $x_{i}$ is the simulated optical depth,  $\mu_{i}$ is the observed optical depth, and $\sigma_{i}$ is the error in the observed optical depth, which we set to 0.1, chosen as a conservative estimate of the source disk from \citet{stark2014}.  The log-likelihood is computed for each pixel over the entire disk. 

Each simulation is run for an amount of time representing the typical collisional lifespan of the particles:  we examined lifespans of 3~kyr to 20~kyr (see Section~\ref{sec:lifetimes}).  In our model we assume that the particles observed in reflected light are created continuously from an underlying population of larger planetesimals.  To represent the continual evolution of these particles from creation (at 0~kyr) to collisional destruction, \cf{101} snapshots (including $t$=0~kyr) taken at equal intervals of time of the disk's evolution are stacked together.  All data is plotted in the planet-centric frame.  In this way, a steady-state disk is achieved which includes particles of all ages from creation to destruction.  However, for simplicity we do not model the time-dependence of the collisional removal, and all particles of all ages count for the same weight in our optical depth maps.  The entire disk is then rotated to the best possible alignment with the Target Map.

For our model, MultiNest typically requires 4000 to 6000 iterations to produce the posteriors.  A typical single MultiNest search takes 3-4 days to complete running in parallel on a machine with 48 Intel(R) Xeon Silver 4214 CPUs. Owing to the lengthy nature of the search we chose not to add additional parameters to the MultiNest search.  Since our purpose is only to show that a planet can plausibly explain the structure seen in HD~181327 and not necessarily to determine that planet's parameters with high precision, this approach is sufficient to achieve our goal.

\section{Results}
\label{sec:results}

\subsection{Initial Observations and Particle Segmentation}
\label{sec:initobs}
In our first set of simulations we examined a range of particles with $\beta$~=~0.05 to 0.8.  As part of our initial analysis, we divided the particles into different $\beta$ ranges to see how behaviour differed.  We found that only a subset of particles, those with $\beta$$\sim$0.5, were able to quickly produce an arc.  If $\beta$ were significantly higher ($\beta$$\ge$0.6), the particles are too easily lost.  Conversely, the larger particles ($\beta$<0.2) were found not to be driven into an arc of high density (for the same planetary parameters).  As discussed in Section~$\ref{sec:partdist}$ particles with $\beta$$\approx$0.5, despite being at or above the notional blowout limit, can persist in the disk in stable orbits depending on their relative ejection speed from their source body.  We therefore focused our efforts on a disk of particles where each particle had a randomly determined $\beta$ value between 0.4 and 0.6 distributed uniformly.

\subsection{Best-Fit Results}
\label{sec:bestfits}
We found that models with 0.4<$\beta$<0.6 and a range of collisional timescales were able to closely approximate the disk observed by \citet{stark2014}.  The posteriors and best-fit results for each timescale can be seen in Table~\ref{tab:bestResults} and figures for some representative cases are included below.  The MultiNest results indicate a complicated parameter landscape, and different sets of coupled parameters (usually mass and semimajor axis) can produce similar quality results.  Several combinations are presented below for comparison.  Our single best-fit was for the 6~kyr timescale, but the 4~kyr and 8~kyr lifetimes are close, each with their own quirks that make them slightly worse fits than the 6~kyr.

We found there were two broad classes of solutions for all timescales.  The first requires a planet of M$_{J}$$\approx$2-4 with a semimajor axis of 68-72~au.  The other requires a more massive planet of 7-12 $M_{J}$ but with a semimajor axis between 50 and 55~au.  For particle lifetimes of less than 10~kyr, the smaller planet solution is the better fit, producing usually better log-likelihood values.  Conversely, for particle lifetimes of 10~kyr or more, the more massive planet solutions produced better fits.  However, \citet{wahhaj2013} placed a lower mass detection limit of 3.5 $\rm M_{J}$.  For this reason, we concentrated our efforts on lower-mass solutions (see Section~\ref{sec:detectlimit}).

Most of our simulations were able to match the extended disk very well, as indicated by low $\chi^{2}$ values beyond 100~au.  Similarly, we are able to reproduce the Target Disk's region between 60 and 100~au.  The width and location of the parent ring in our simulations are similar to the Target's, though the $\chi^{2}$ values shown there is more variation within it than in the extended cloud.  Our starting annulus, which we model as the source of particles, is typical centered at 90~au and between 15 and 20~au in width (see disk width and center posteriors in Table~\ref{tab:bestResults}).  It is only through the evolution, the continual creation and motion through the collisional lifetime, of particles that we see the most dense region form at 84.2~au.

The most sensitive regions were the arc itself and the inner edge of the disk.  No result was able to produce an arc with the same intensity as the Target Disk: our best fits in the western arc have an average value 7.4\% higher than the average of the other quadrants, compared to 12\% increase seen in the Target Disk.  Overall, our simulations are able to match the overall pattern with a 90$^\circ$ arc similar to the Target Disk.

Below we briefly describe the best-fit result for each timescale (corresponding to the assumed collisional lifetime of the particles), and include the simulated disk and $\chi^{2}$ per pixel maps.  The location of the planet is indicated by a white $\times$, and all maps are in the planet-centric frame.

\begin{table*}
	\centering
	\caption{Posteriors and Best-Fit Results at Different Particle Lifetimes. The three $\chi^2$ per pixel entries correspond to specific portions of the disk: entire disk from 60 to 140 au, the parent ring refers to the annulus from 70~au to 100~au, and the arc refers 70~au and 100 au of the upper-right (western) quadrant only. }
	\label{tab:bestResults}
	\begin{tabular}{rccccccc} 
	       & 3~kyr & 4~kyr & 6~kyr & 8~kyr & 10~kyr & 14~kyr & 20~kyr \\  
		\hline
		Posterior mass ($\rm M_{J}$) & 3.7$\substack{+4.7\\-2.1}$ & 3.6$\substack{+2.7\\-2.5}$ & 3.1$\substack{+2.6\\-2.1}$ & 2.7$\substack{+1.1\\-1.8}$ & 2.0$\substack{+1.7\\-1.3}$ & 4.1$\substack{+2.6\\-2.8}$ & 2.4$\substack{+2.3\\-1.6}$ \\ [5pt]
		Posterior semimajor axis (au) & 71.6$\substack{+3.1\\-4.1}$ & 71.8$\substack{+2.7\\-3.2}$ & 71.3$\substack{+3.5\\-1.9}$ & 71.5$\substack{+2.8\\-2.0}$ & 70.8$\substack{+3.1\\-2.1}$ & 70.8$\substack{+3.8\\-2.9}$ & 71.5$\substack{+3.5\\-2.7}$ \\ [5pt]
        Posterior parent width (au) & 16.9$\substack{+4.3\\-3.6}$ & 18.6$\substack{+5.6\\-4.0}$ & 17.6$\substack{+6.6\\-3.8}$ & 17.7$\substack{+4.0\\-4.5}$ & 19.8$\substack{+3.1\\-4.5}$ & 17.7$\substack{+4.1\\-3.0}$ & 18.5$\substack{+6.8\\-4.1}$ \\ [5pt]
        Posterior parent center (au) & 89.2$\substack{+1.0\\-0.9}$ & 90.2$\substack{+1.5\\-1.6}$ & 89.8$\substack{+1.9\\-0.9}$ & 89.9$\substack{+1.4\\-1.5}$ & 90.3$\substack{+2.1\\-1.3}$ & 89.5$\substack{+1.3\\-1.2}$ & 89.9$\substack{+2.3\\-1.3}$ \\ [5pt]
        Posterior ejection speed (m/s) & 1273$\substack{+22\\-20}$ & 1261$\substack{+26\\-25}$ & 1265$\substack{+8\\-12}$ & 1258$\substack{+28\\-15}$ & 1260$\substack{+14\\-14}$ & 1276$\substack{+23\\-26}$ & 1273$\substack{+21\\-25}$ \\ [5pt]
        Best fit mass ($\rm M_{J}$) & 2.65 & 2.44 & 3.64 & 3.65 & 3.71 & 4.62 & 3.83 \\ [5pt]
		Best fit semimajor axis (au) & 72.82 & 73.61 & 70.15 & 69.48 & 68.72 & 53.76 & 69.38 \\ [5pt]
        Best fit parent width (au) & 21.26 & 19.83 & 13.86 & 13.22 & 15.27 & 21.88 & 20.15 \\ [5pt]
        Best fit parent center (au) & 88.99 & 90.27 & 88.94 & 88.34 & 89.00 & 91.73 & 88.63 \\ [5pt]
        Best fit ejection speed (m/s) & 1263.5 & 1251.5 & 1265.8 & 1263.2 & 1274.2 & 1264.6 & 1275.6 \\ [5pt] 
        $\chi^{2}$ per pixel of entire disk & 0.144 & 0.102 & 0.091 & 0.094 & 0.092 & 0.119 & 0.098 \\ [5pt]
        $\chi^{2}$ per pixel of parent ring & 0.288 & 0.189 & 0.142 & 0.143 & 0.148 & 0.210 & 0.175 \\ [5pt]
        $\chi^{2}$ per pixel of $90^\circ$ of arc & 0.302 & 0.227 & 0.175 & 0.160 & 0.208 & 0.223 & 0.225 \\ [5pt]
	\end{tabular}
\end{table*}

\subsubsection{Collisional Timescale of 3~kyr}
\begin{figure}
 \includegraphics[scale=0.35]{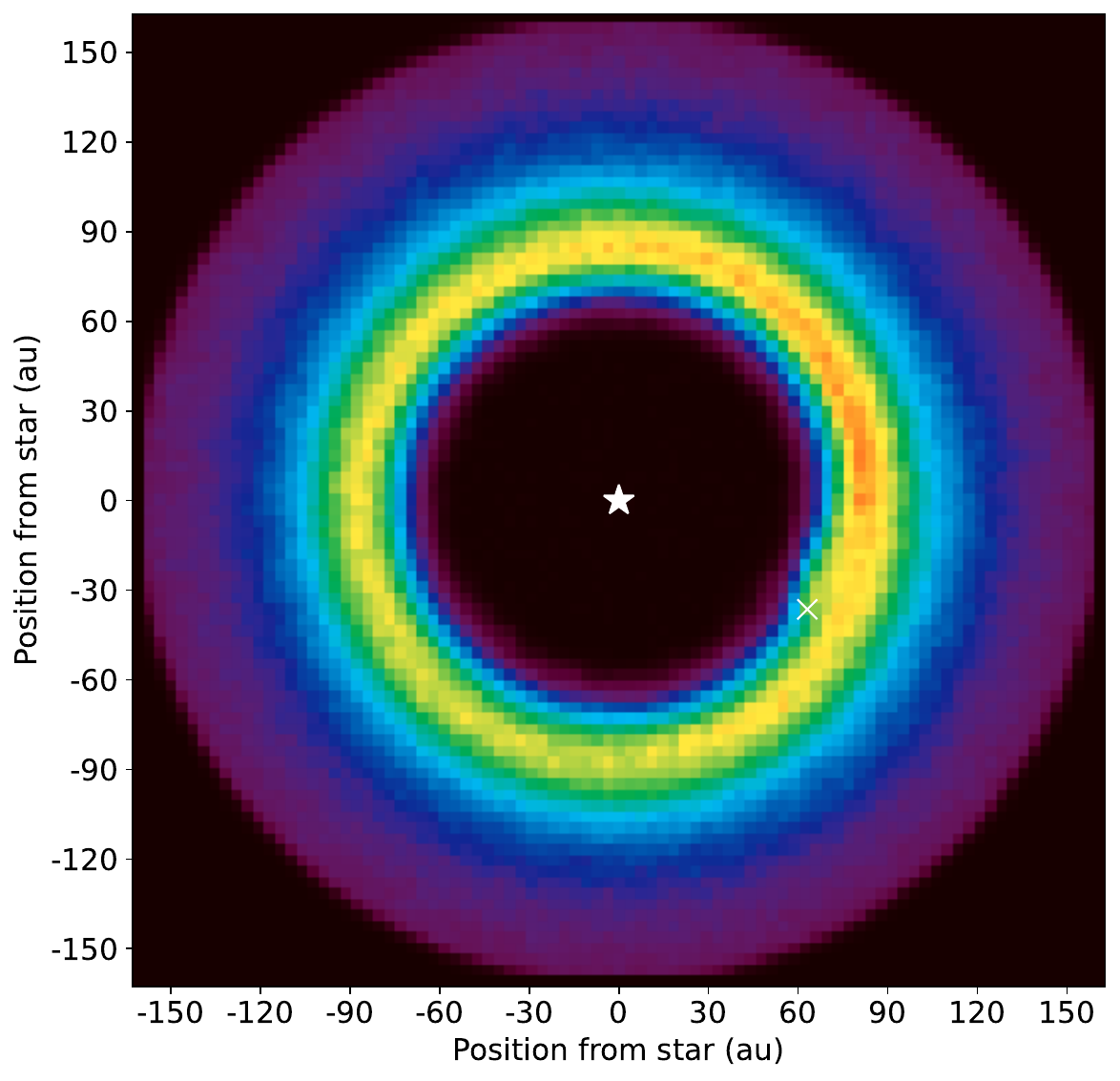}
 \includegraphics[scale=0.51, trim=0 -30 -20 0]{stark_scale_vertical_narrow.pdf}
 \includegraphics[scale=0.35]{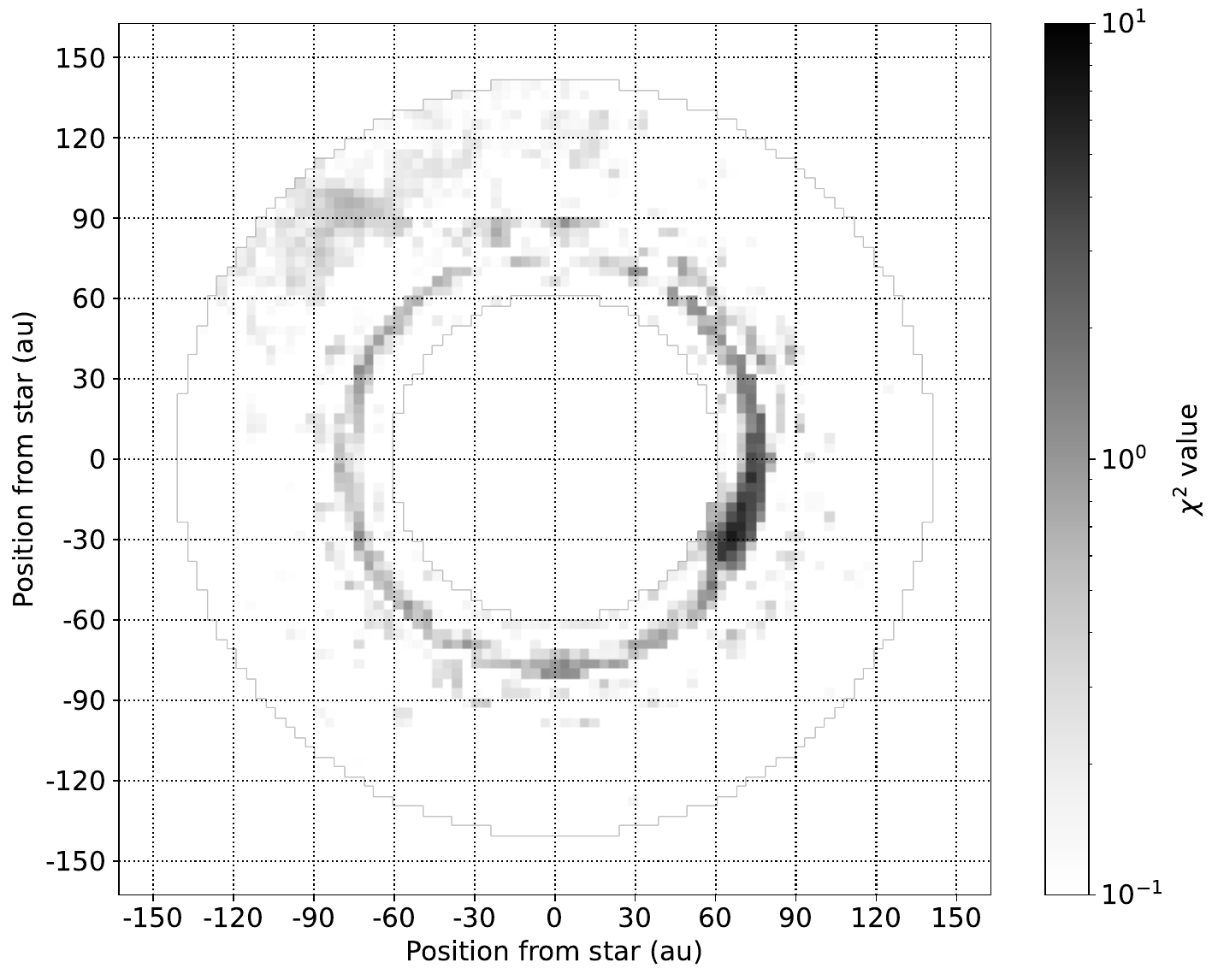}
 \caption{\label{fig:run74m}Best Fit result for a particle lifespan of 3~kyr.  The top image is the simulated optical depth map, with the white $\times$ marking the location of the planet.  At the bottom is the $\chi^{2}$ map when compared to the Target Map.  Both the planet and particles orbit in the clockwise direction.}
\end{figure}

The shortest timescale we simulated was 3~kyr, translating to only six orbits of the proposed planet.  The 2.6~$\rm M_{J}$ planet at 72.8~au creates a distinct $\sim$70$^\circ$ arc, the most intense of any results (8.6\% compared to the Target Disk's 12.5\%).  However there are significant mismatches on the inner edge, particularly in the south-west near the planet.  As a result, the overall $\chi^{2}$ values were the highest in all regions.  Despite the high $\chi^{2}$ values, the intensity of the arc makes this a viable solution.  The results are shown in Figure~\ref{fig:run74m}.

\subsubsection{Collisional Timescale of 4~kyr}
\begin{figure}
 \includegraphics[scale=0.35]{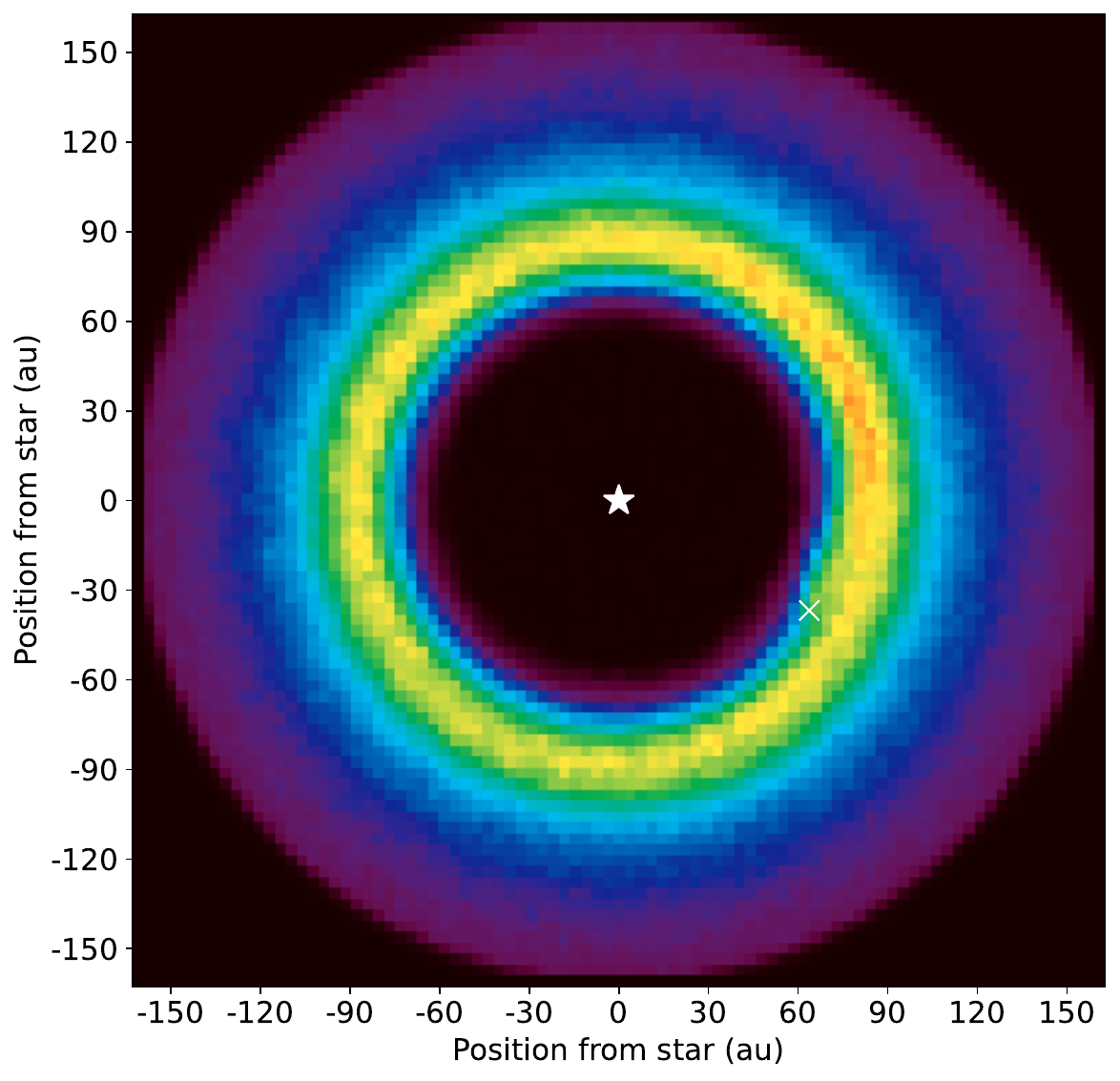}
 \includegraphics[scale=0.51, trim=0 -30 -20 0]{stark_scale_vertical_narrow.pdf}
 \includegraphics[scale=0.35]{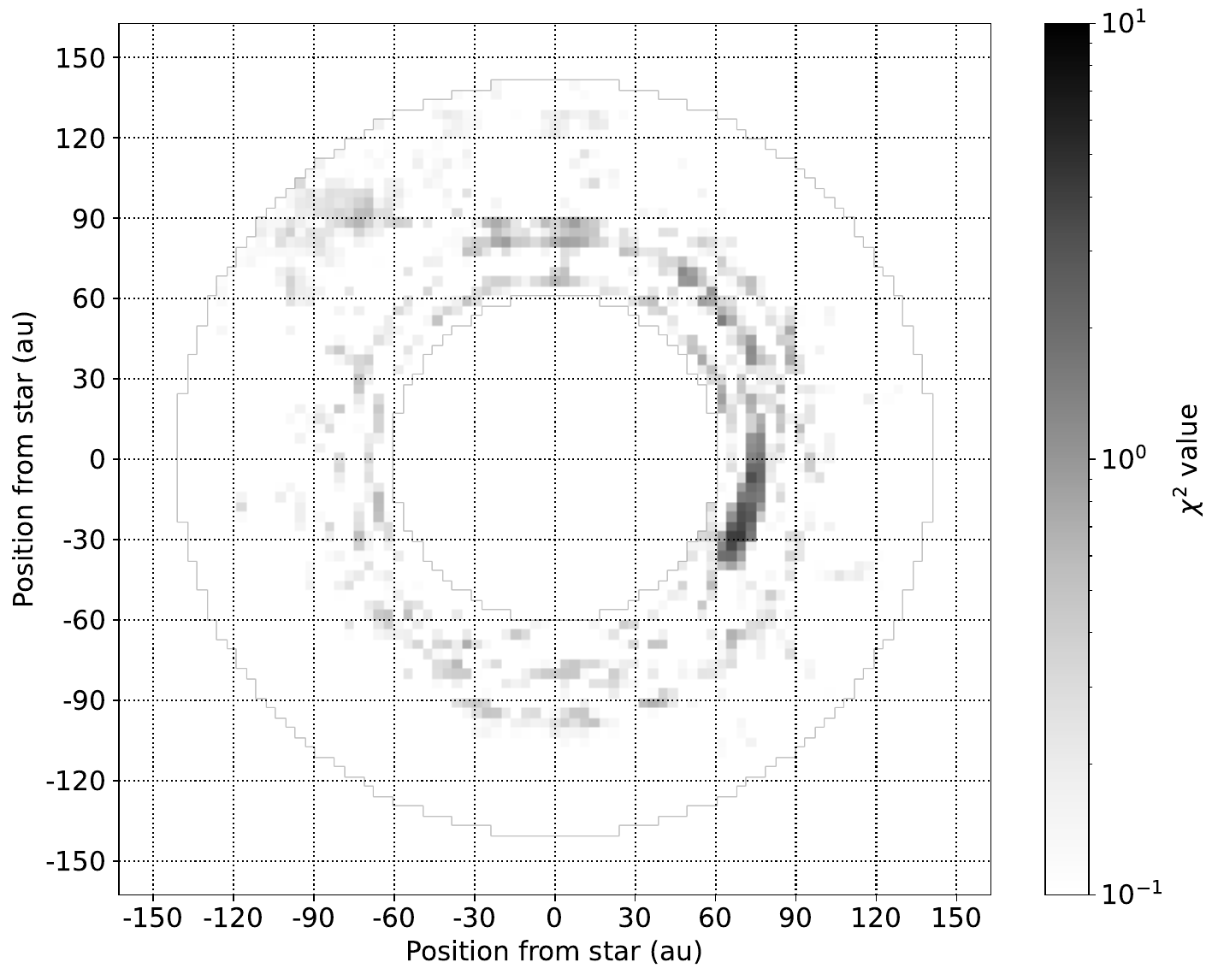}
 \caption{\label{fig:run74i}Best Fit result for a particle lifespan of 4~kyr.  Panels and symbols are identical to those described in Figure~2.}
\end{figure}

The 4~kyr lifetime translates to about eight orbits of the proposed planet.  The planet (2.4~$\rm M_{J}$ at 73.6~$\rm au$) is able to create a short arc while removing the larger mismatches seen at the 3~kyr result.  A smaller tadpole-shaped region  on the inner edge remains, but is significantly reduced.  This keeps the $\chi^{2}$ of the arc region relatively high.  The results are shown in Figure~\ref{fig:run74i}.

\subsubsection{Collisional Timescale of 6~kyr}
\begin{figure}
 \includegraphics[scale=0.35]{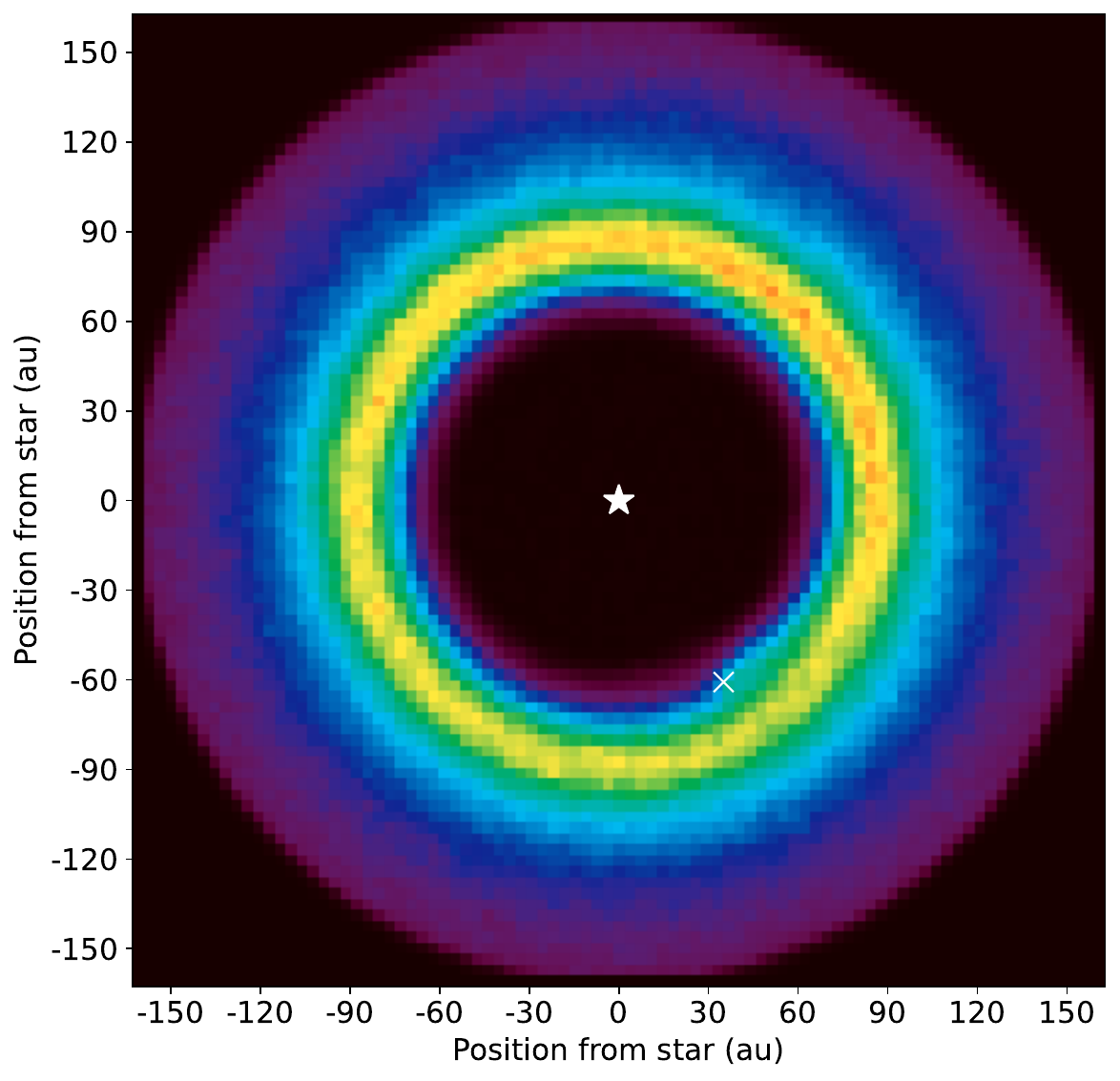}
 \includegraphics[scale=0.51, trim=0 -30 -20 0]{stark_scale_vertical_narrow.pdf}
 \includegraphics[scale=0.35]{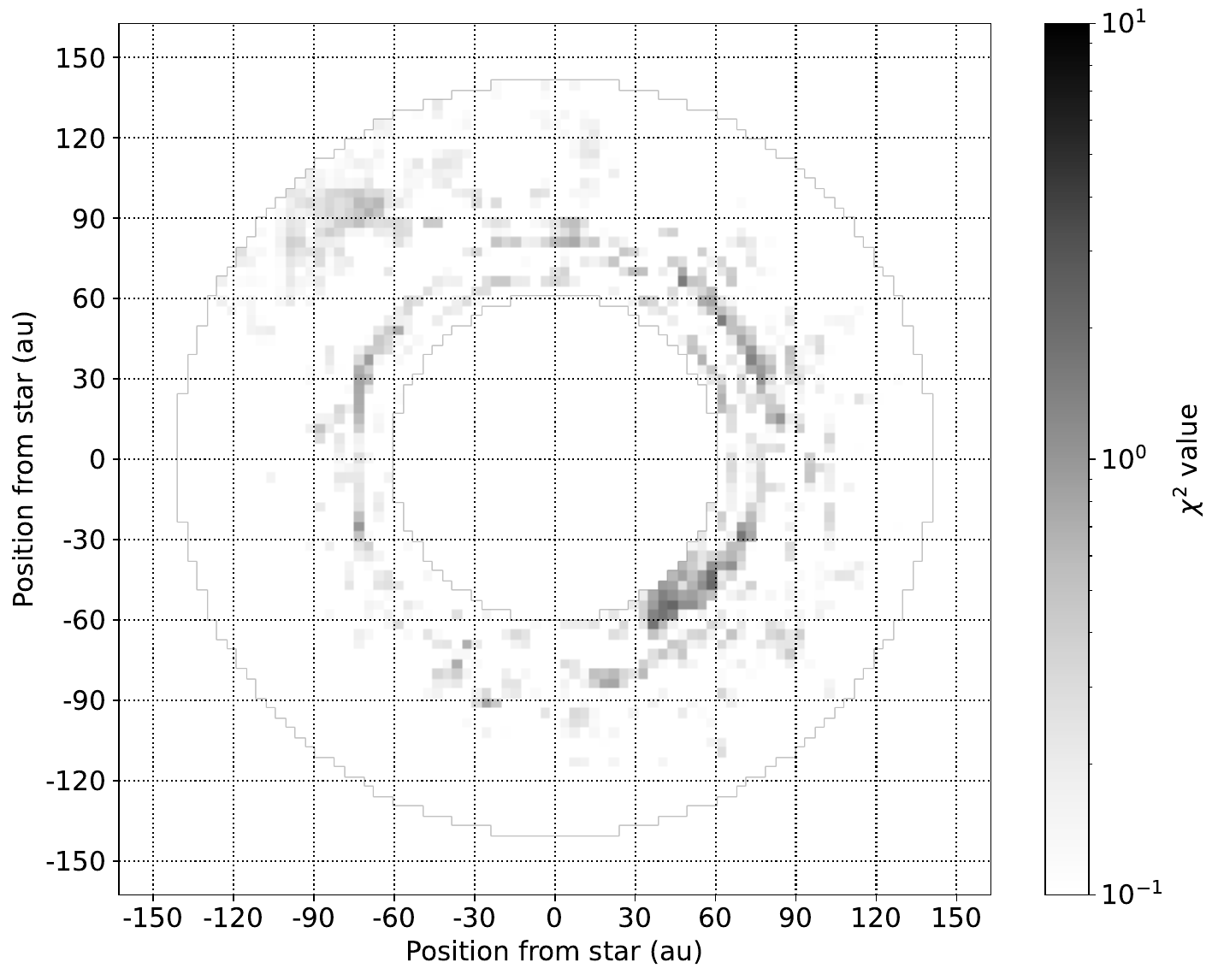}
 \caption{\label{fig:run74g}Best Fit result for a particle lifespan of 6~kyr.  Panels and symbols are identical to those described in Figure~2.}
\end{figure}

Our overall best fit occurs at the 6~kyr timescale.  A planet of 3.6~$\rm M_{J}$ at 70.1~$\rm au$ creates an arc of 90$^\circ$.  The arc is not as intense as our Target Map, with an arc that exceeds the parent average by 7.5\%, compared to 12\% of the Target Disk.  This represents our single best overall result of this study, producing an arc of 90$^\circ$, with the lowest overall $\chi^{2}$ values.  The results are shown in Figure~\ref{fig:run74g}.

\subsubsection{Collisional Timescale of 8~kyr}
\begin{figure}
 \includegraphics[scale=0.35]{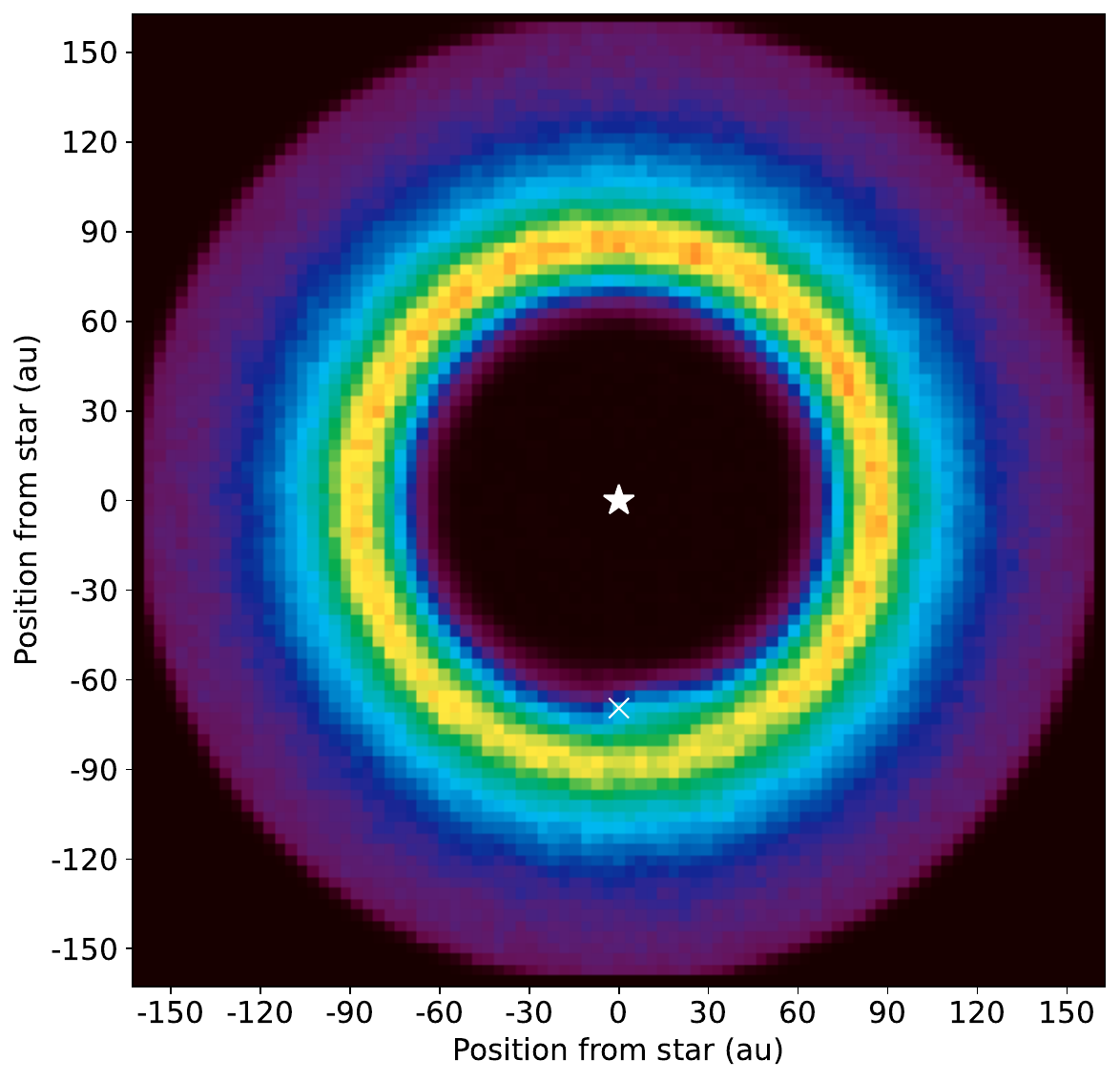}
 \includegraphics[scale=0.51, trim=0 -30 -20 0]{stark_scale_vertical_narrow.pdf}
 \includegraphics[scale=0.35]{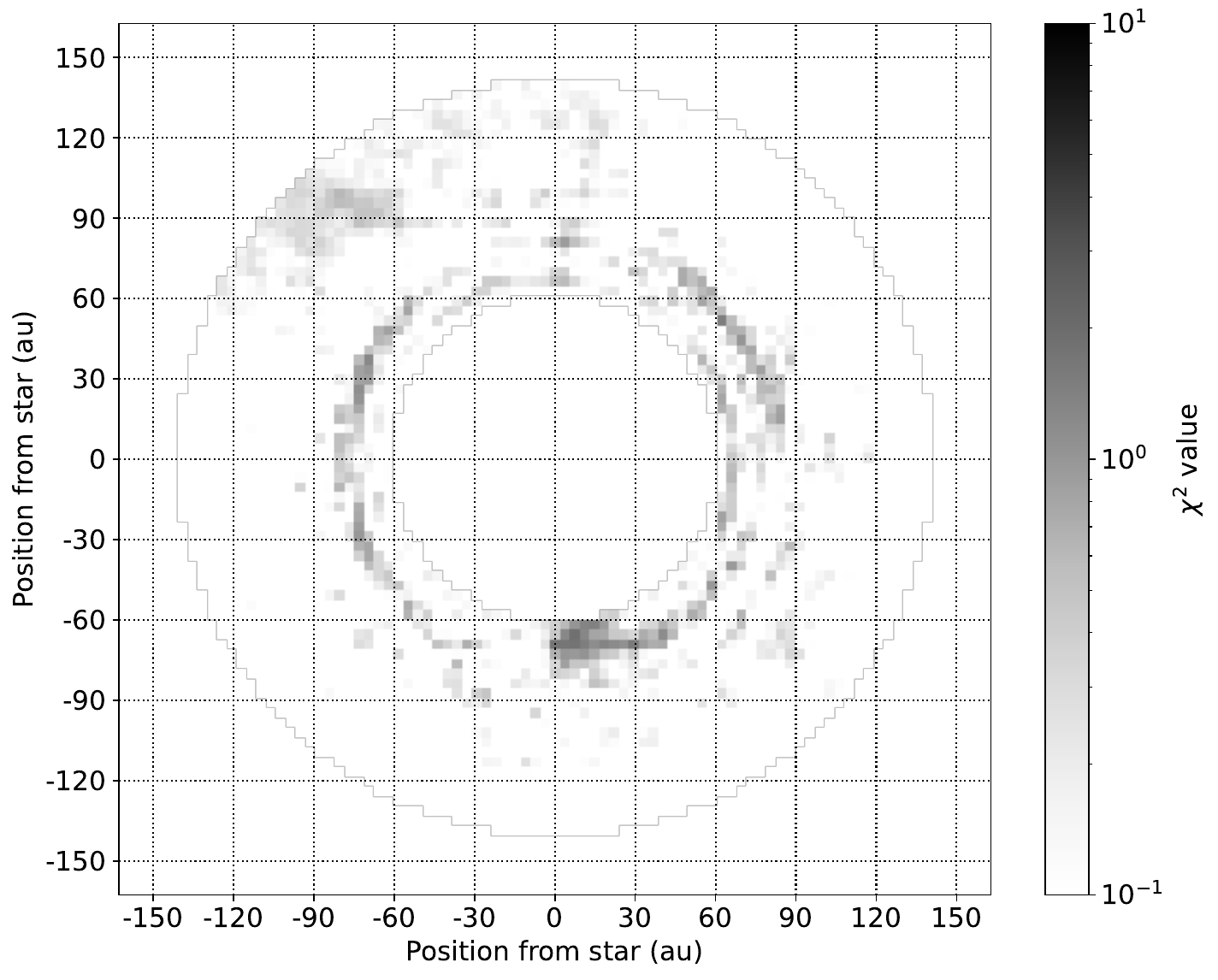}
 \caption{\label{fig:run74j}Best Fit result for a particle lifespan of 8~kyr.  Panels and symbols are identical to those described in Figure~2.}
\end{figure}

The 8~kyr timescale produces an oversized arc, almost 180$^\circ$ in expanse.  It does a good job at matching the arc itself, the best of any result.  However the arc is 180$^\circ$ extending into the northern quadrant.  Due to a dearth in the southern side, the average value and overall $\chi^{2}$ of the parent ring is among our best.  The north-south dichotomy prevents this from being our best result.  This planet has a mass of 3.6~$\rm M_{J}$ and semimajor axis of 69.5~$\rm au$.  The results are shown in Figure~\ref{fig:run74j}.

\subsubsection{Collisional Timescale of 10~kyr}
\begin{figure}
 \includegraphics[scale=0.35]{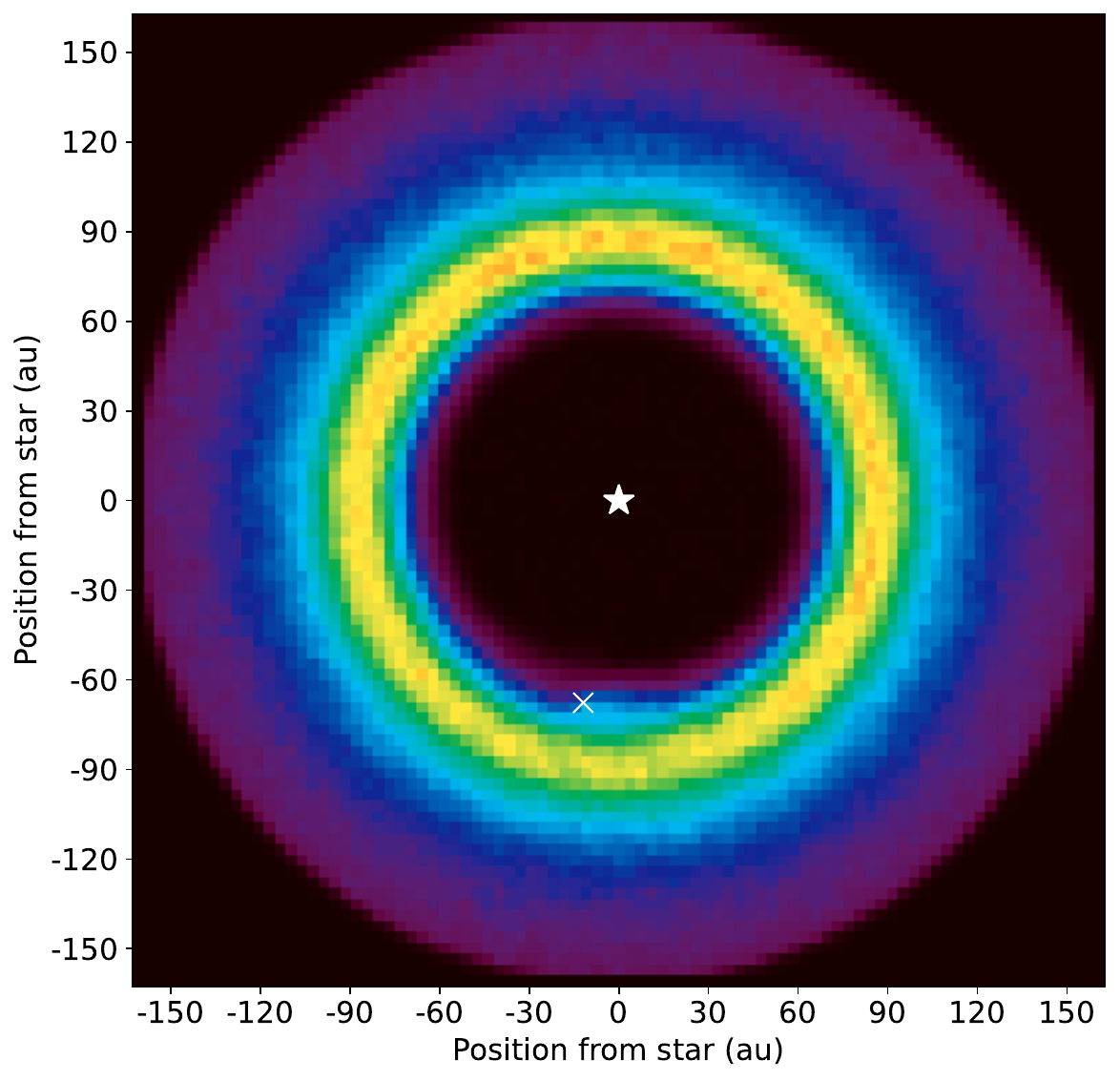}
 \includegraphics[scale=0.51, trim=0 -30 -20 0]{stark_scale_vertical_narrow.pdf}
 \includegraphics[scale=0.35]{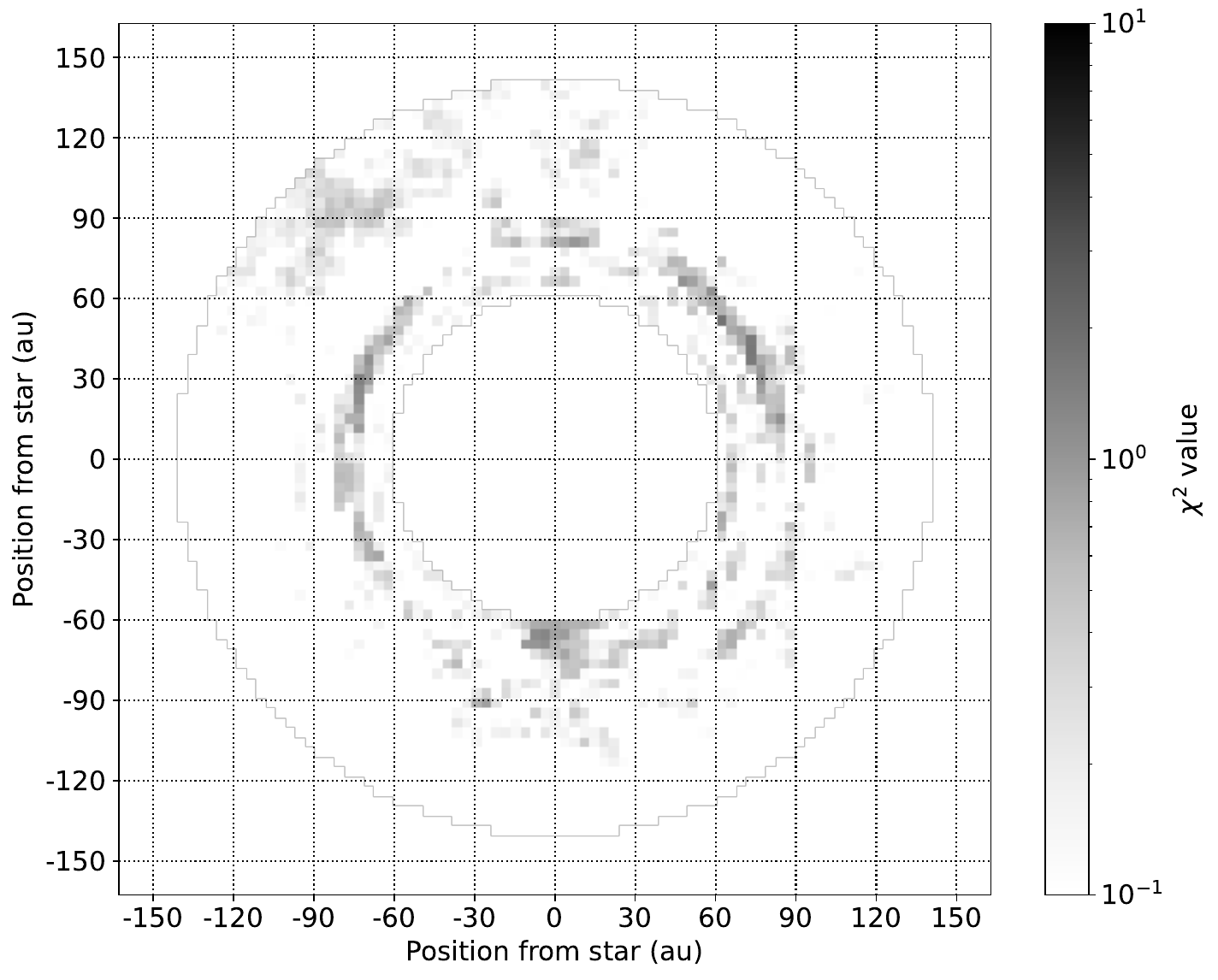}
 \caption{\label{fig:run74h}Best Fit result for a particle lifespan of 10~kyr.  Panels and symbols are identical to those described in Figure~2.}
\end{figure}

The 10~kyr timescale produces an inconsistent arc of about 120$^\circ$.  The arc is not well-defined and not very intense.  Here, the planet has a mass of 3.7~$\rm M_{J}$ and semimajor axis of 68.7~au.  The results are shown in Figure~\ref{fig:run74h}.

\subsubsection{Collisional Timescale of 14~kyr}
\begin{figure}
 \includegraphics[scale=0.35]{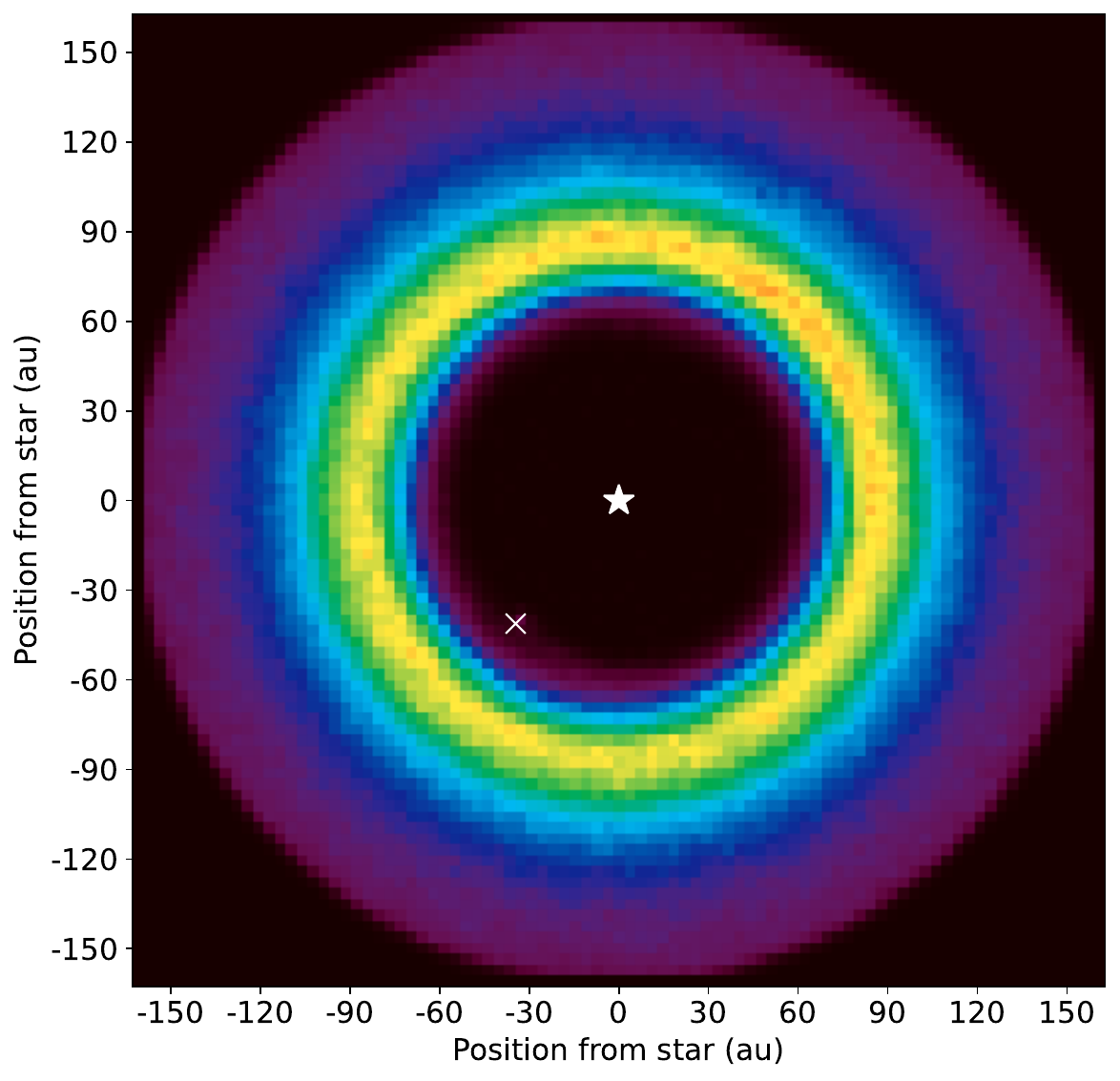}
 \includegraphics[scale=0.51, trim=0 -30 -20 0]{stark_scale_vertical_narrow.pdf}
 \includegraphics[scale=0.35]{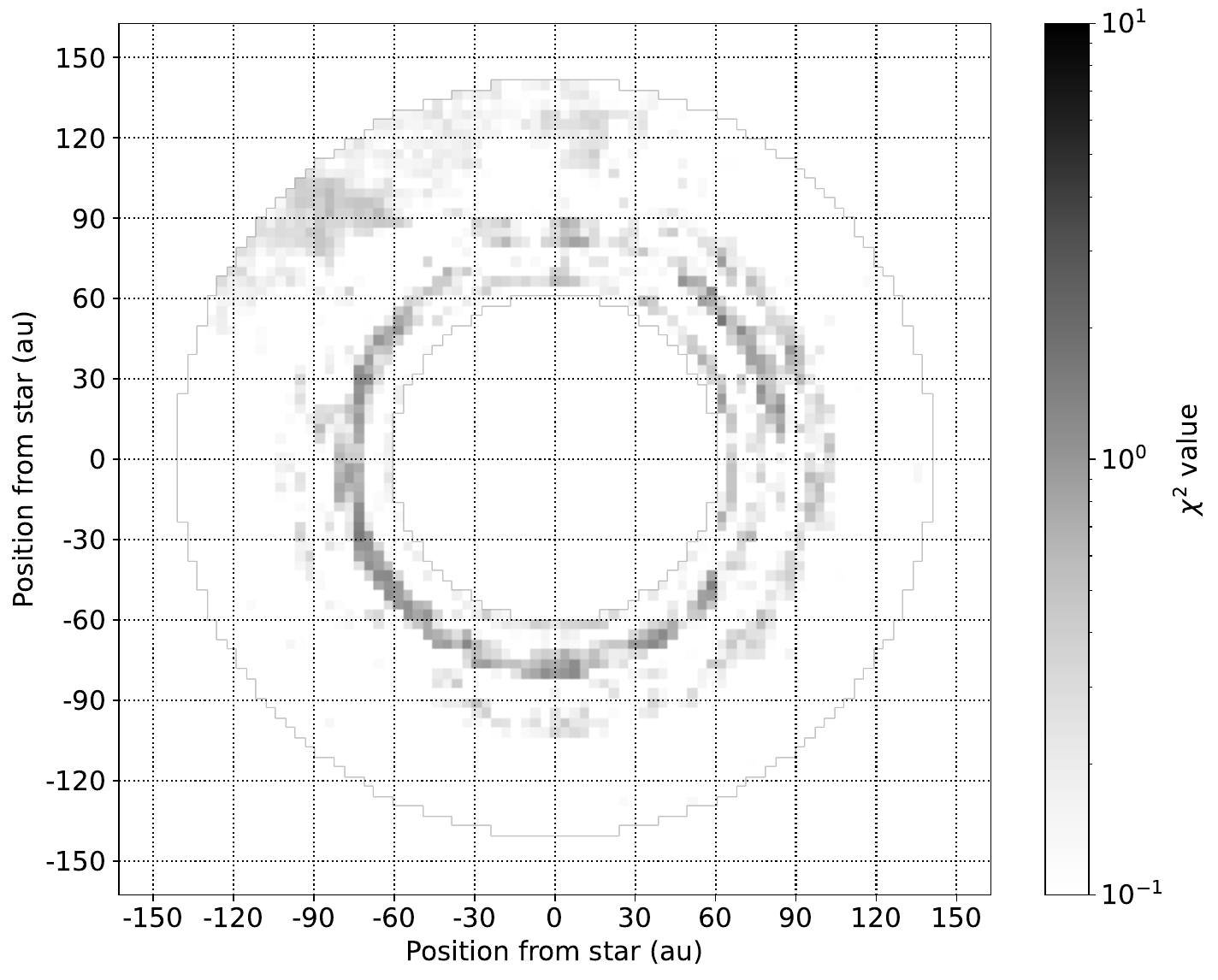}
 \caption{\label{fig:run74k}Best Fit result for a particle lifespan of 14~kyr.  Panels and symbols are identical to those described in Figure~2.}
\end{figure}

The best solution at the 14~kyr timescale required a 11~$\rm M_{J}$ planet.  As this is well above the limit set by \citet{wahhaj2013} (see Section~\ref{sec:detectlimit}) we instead looked at lower-mass solutions.  A 4.6~$\rm M_{J}$ at 53.8~au produces a thin 90$^\circ$ arc.  The result is similar to the 10~kyr result.  Though here, the eastern side of the disk is not fit as well as in other lifetimes.  Overall, this has the second-worst $\chi^{2}$ values of all our examined lifetimes.  The results are shown in Figure~\ref{fig:run74k}.

\subsubsection{Collisional Timescale of 20~kyr}
\begin{figure}
 \includegraphics[scale=0.35]{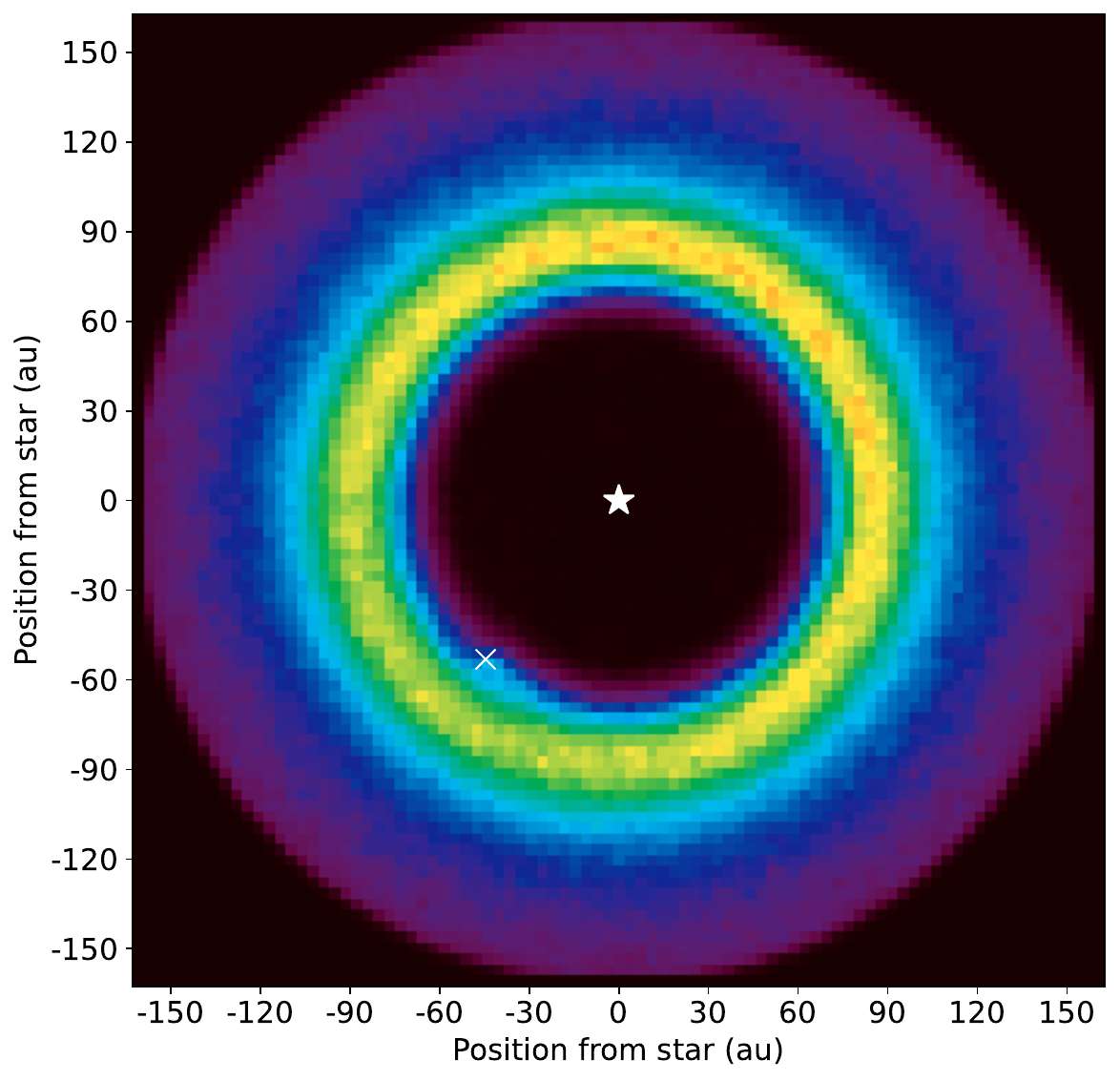}
 \includegraphics[scale=0.51, trim=0 -30 -20 0]{stark_scale_vertical_narrow.pdf}
 \includegraphics[scale=0.35]{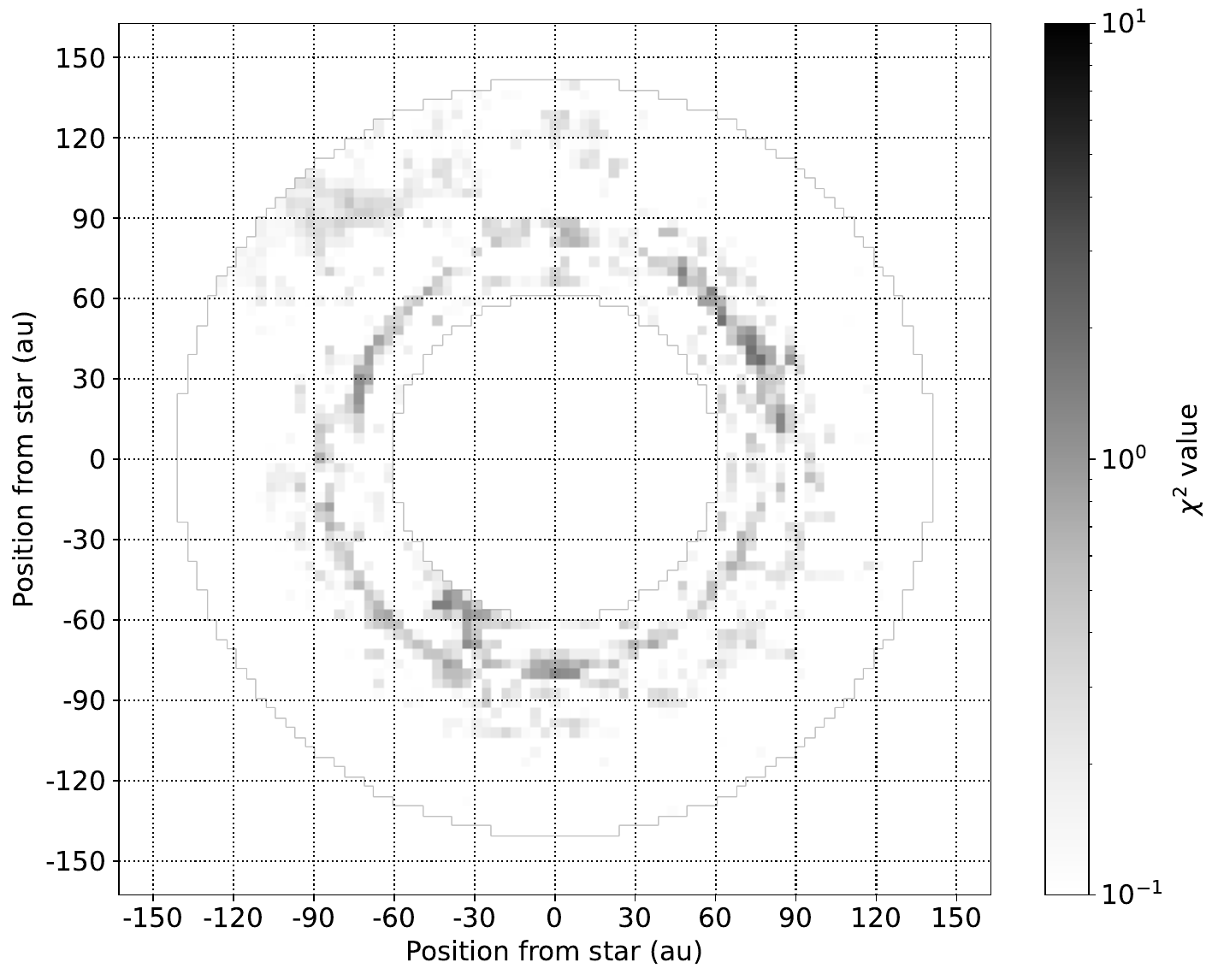}
 \caption{\label{fig:run74f}Best Fit result for a particle lifespan of 20~kyr.  Panels and symbols are identical to those described in Figure~2.}
\end{figure}

Our longest lifetime was 20~kyr.  Like the 14~kyr run, the overall best result required an 11~$\rm M_{J}$ planet, which we rule out \citep{wahhaj2013}.  Our best low-mass solution consists of a 3.8~$\rm M_{J}$ planet with a semimajor axis of 69.4~$\rm au$.  The result is a spotty 90$^\circ$ arc, and the far eastern side has a dearth of particles.  The relative intensity of the arc (6.5\% greater than the rest of the parent disk) appears comparable to our best runs, but that is due to the low density on the eastern side.  The results are shown in Figure~\ref{fig:run74f}.

\subsection{Relative Particle Density of Arc to Parent Ring}
To quantify the quality of our simulations, in particular the strength of the arc, we analyzed and compared the particle counts in the regions of interest- the parent disk and the arc- between our simulations and the Target Disk.

\begin{table*}
	\centering
	\caption{Average Optical Depth per Pixel in Each Quadrant of the Parent Disk}
	\label{tab:relparticles}
	\begin{tabular}{rcccccccc} 
	     Quadrant & Target Disk & 3~kyr & 4~kyr & 6~kyr & 8~kyr & 10~kyr & 14~kyr & 20~kyr \\  
		\hline
		West & 0.738 & 0.735 & 0.709 & 0.712 & 0.711 & 0.703 & 0.704 & 0.700 \\ [5pt]
		North & 0.681 & 0.680 & 0.665 & 0.686 & 0.700 & 0.695 & 0.685 & 0.679 \\ [5pt]
        East & 0.641 & 0.652 & 0.639 & 0.644 & 0.661 & 0.653 & 0.673 & 0.622 \\ [5pt]
        South & 0.657 & 0.697 & 0.678 & 0.660 & 0.681 & 0.680 & 0.682 & 0.674 \\ [5pt]
        \hline
        Increase of West arc\\compared to other quadrants & 11.86\% & 8.65\% & 7.35\% & 7.43\% & 4.51\% & 4.04\% & 3.47\% & 6.30\% \\ [5pt]
	\end{tabular}
\end{table*}

We compared the sum of the optical depths between 70~au and 100~au for each quadrant.  The western quadrant (where the arc of increased density resides) is compared to the average value of the other three quadrants.  This provides a measure of how many more particles exist in the arc than not.  For the Target Disk, this value is 11.9\%, whereas our best-fit simulations show an increase of 7.4\% (for the 6~kyr result, see Figure~\ref{fig:run74g}), and the 4~kyr result is very close (Figure~\ref{fig:run74i}).  The 8~kyr timescale shows a lower 4.5\% increase in arc, but the arc itself is a good fit.  This discrepancy is likely due to its arc extending into the southern quadrant thereby raising the average of the parent ring.  The 10~kyr and 14~kyr lifetimes are only 4.0\% and 3.5\% respectively, indicating the arc for those lifetimes are not as well defined.  The 20~kyr result bucks this trend and increases to 6.3\%, but this appears to be due to a dearth of particles on the east side rather than a surplus on the west (see Figure~\ref{fig:run74f}).

Overall, the best fits were for timescales of 4~kyr, 6~kyr, and 8~kyr.  The quality of fit as measured by $\chi^{2}$, and the relative intensity of the arc are both better matches for the shorter timescales.  The 6~kyr result is our best fit, with the 4~kyr producing a similar but slightly shorter arc, and the 8~kyr producing an arc that is too long.  Collisional lifetimes of 10~kyr and longer either do not produce a strong arc, or produce an arc but with a worse fit to the rest of the disk, or require a planet of significantly higher mass which we reject due to reasons in Section~\ref{sec:detectlimit}.

\subsection{Simulated Disk Profile}
\label{diskprofile}
\begin{figure*} 
 \includegraphics[scale=0.25]{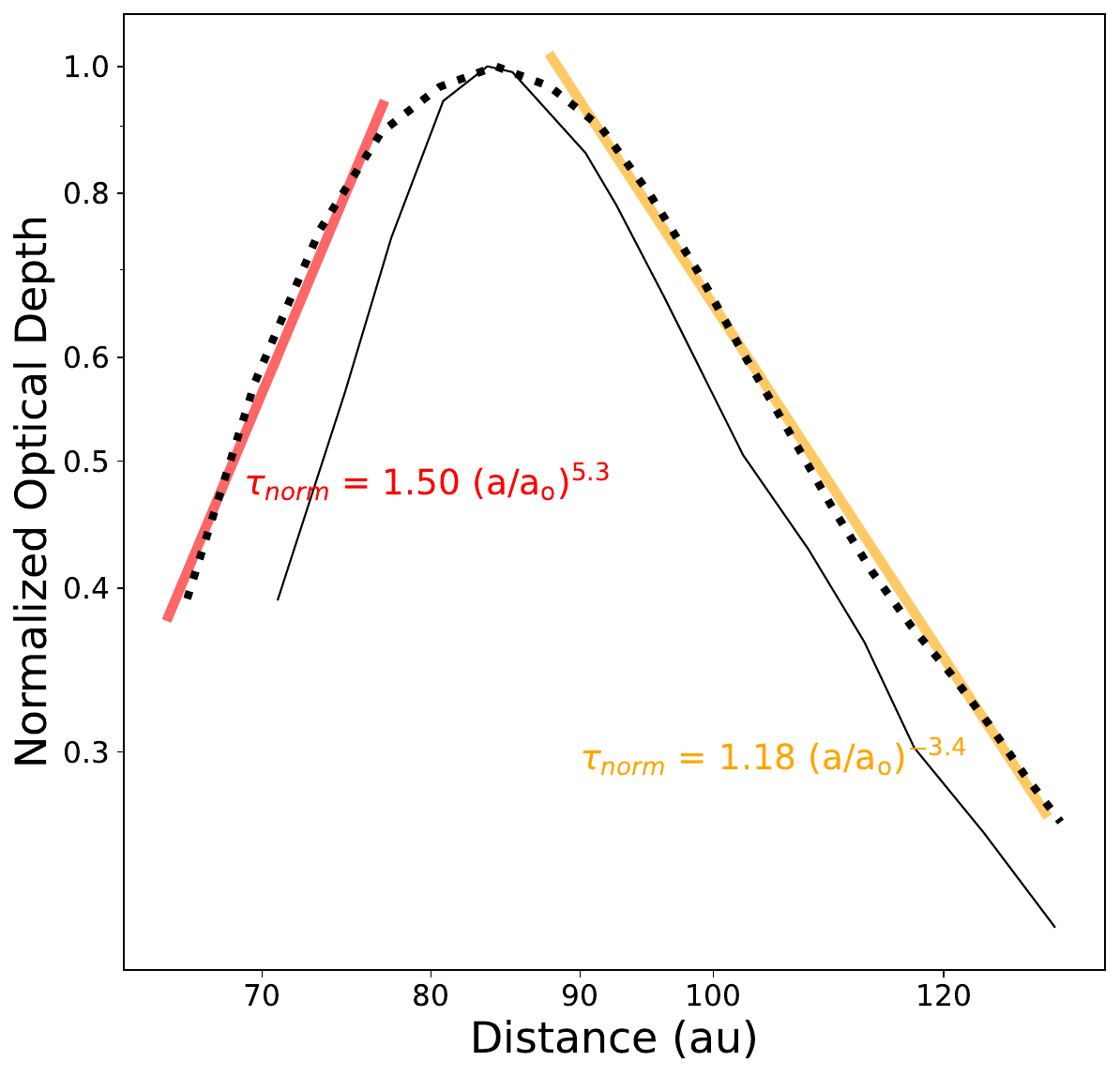}
 \includegraphics[scale=0.25]{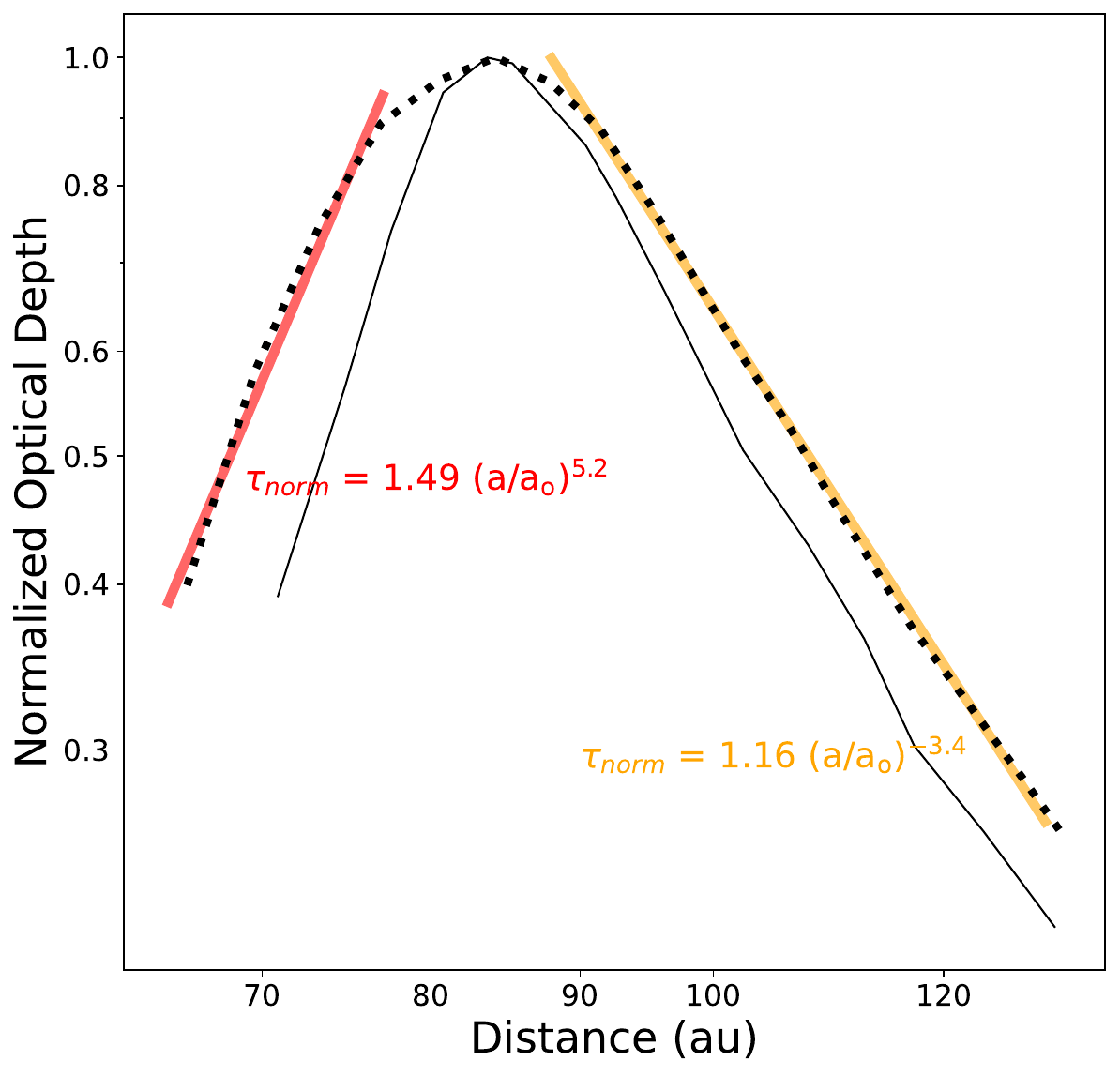}
 \includegraphics[scale=0.25]{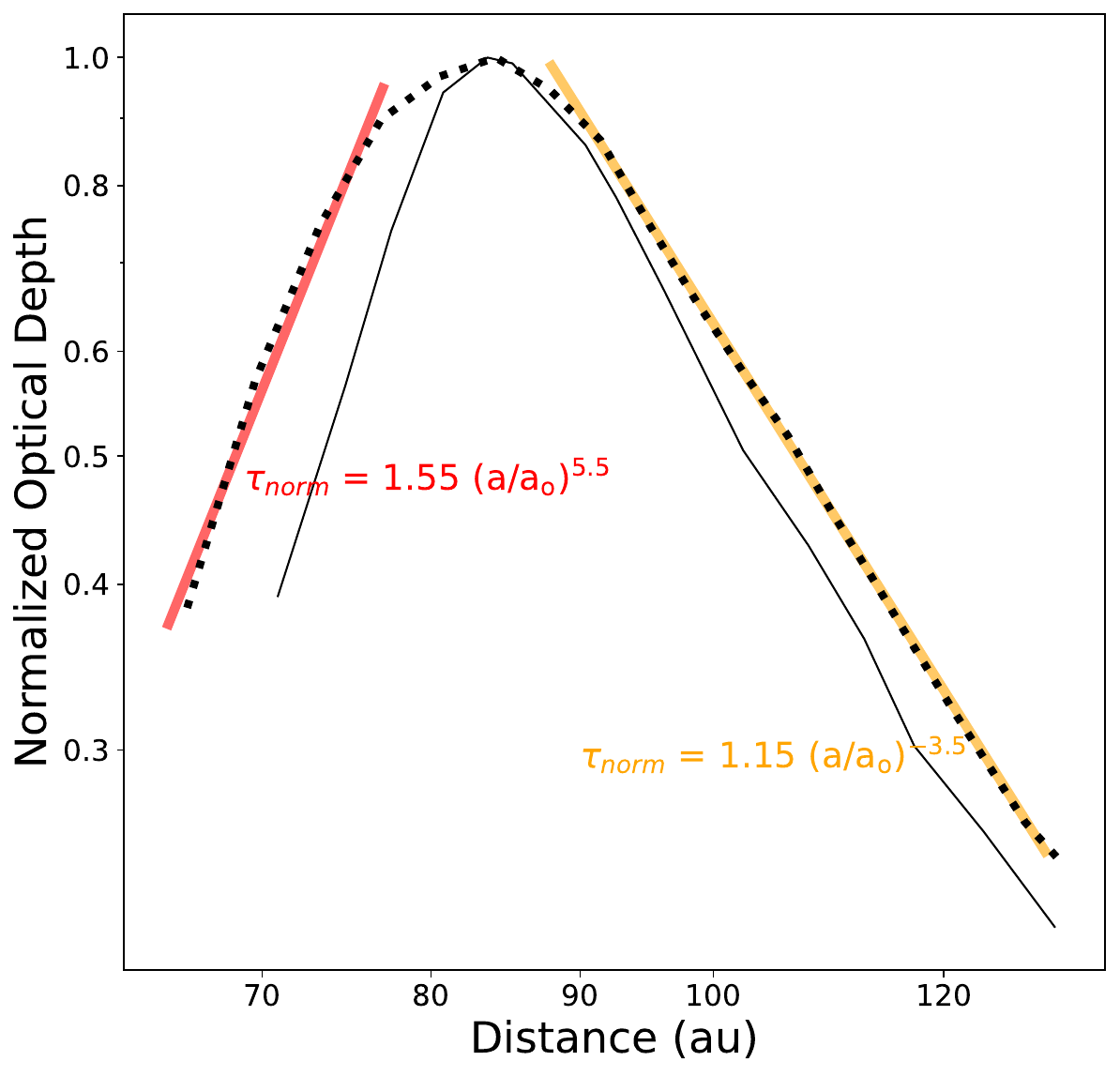}
 \caption{\label{fig:fitProfiles}Normalized mean optical depth profiles of the disks.  On the left is the 4~kyr best-fit result, the middle is the 6~kyr best-fit, and the right is the 8~kyr result.  The dotted line is our simulated result (see Table~\ref{tab:bestResults}), with the best-fit lines for our data shown in red and yellow.  The slopes of the optical depth from 90 to 130~au closely match the slopes found in \citet{stark2014}, which we recreate here with the solid black line.  The inner slope is still steep but less than those from Stark.}
\end{figure*}

To further test our best results, we examined the surface density profiles of the best-fit disks.  Figure~\ref{fig:fitProfiles} shows the normalized (peak value 1) mean optical depth profiles averaged radially over the entire disk.  The inner edge slope is proportional to $a^{5.3}$ while the falloff outside the peak is proportional to $a^{-3.5}$.  Our simulated inner edge is not as sharp as Figure~9 from \citet{stark2014}, with a value of 7.2 (rescaled and re-fitted after adjustment for updated Gaia distance).  However, the outer side of of our disk is better fitted, with our value of -3.4 compared to Stark's -3.8.  The 8~kyr result has a slightly better match on both the inner and outer sides of the disk.

\subsection{Particle Semimajor Axis Distribution Evolution}
\label{sec:resultdist}
\begin{figure*} 
 \includegraphics[scale=0.21]{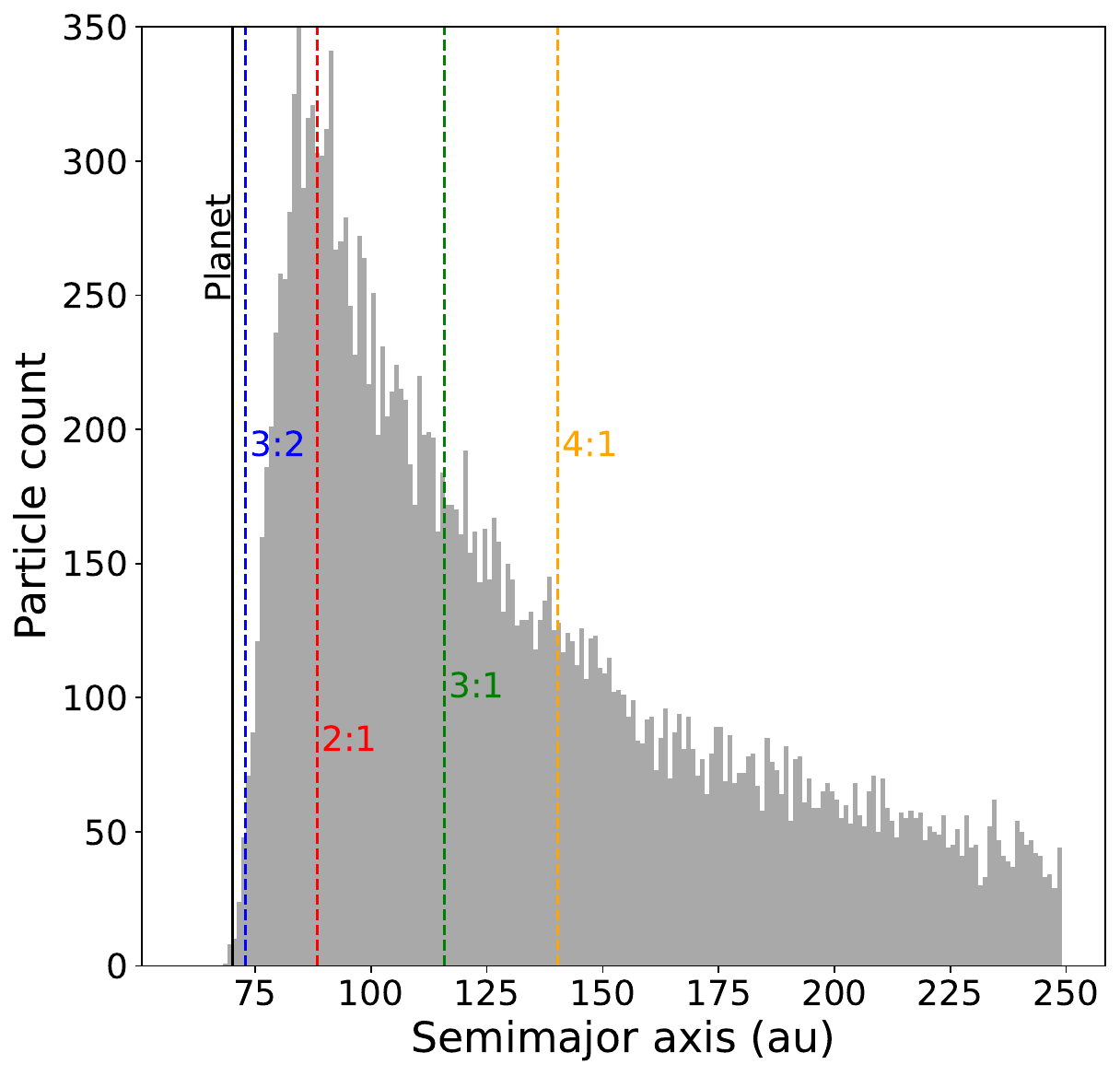}
 \includegraphics[scale=0.21]{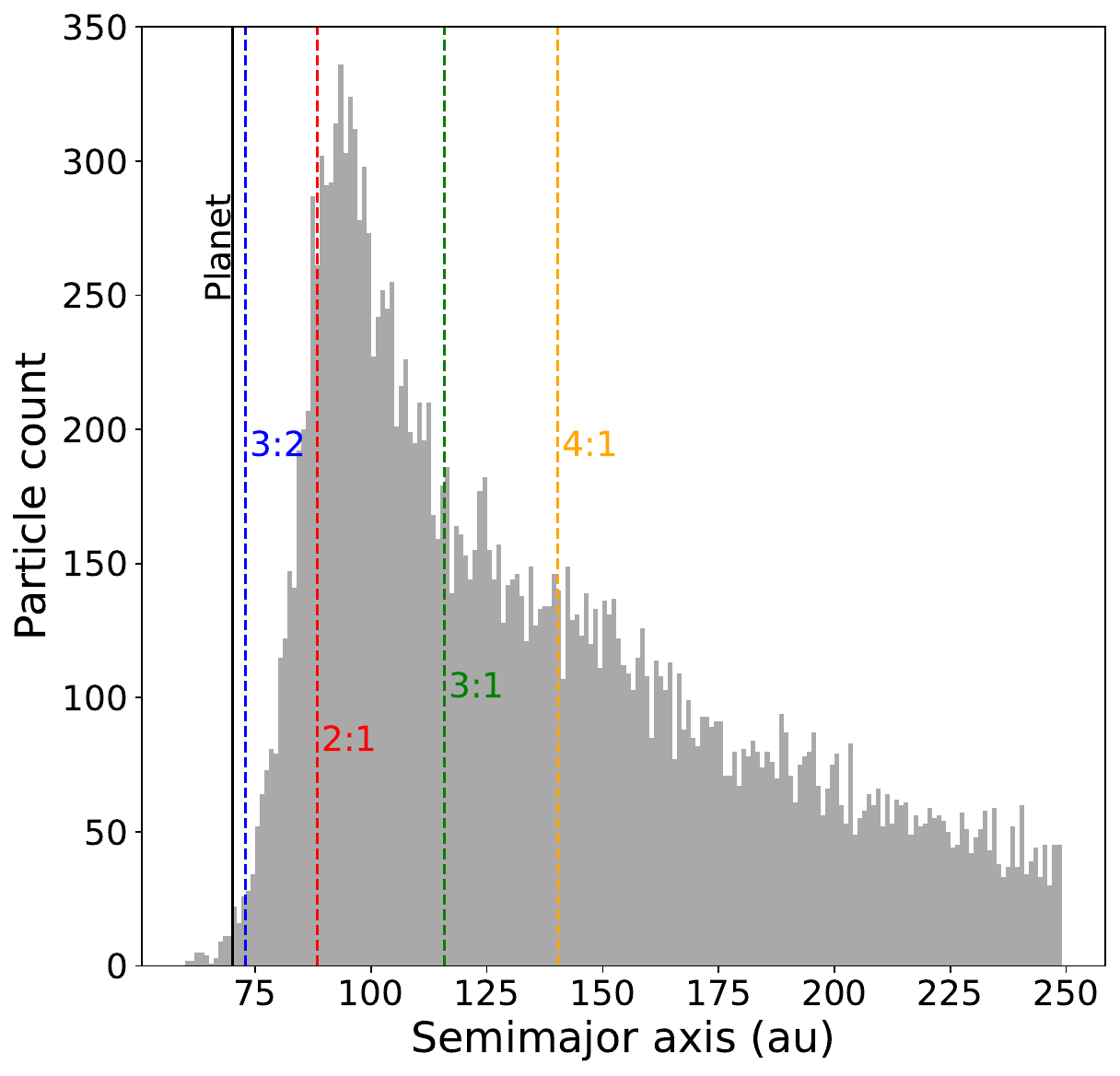}
 \includegraphics[scale=0.21]{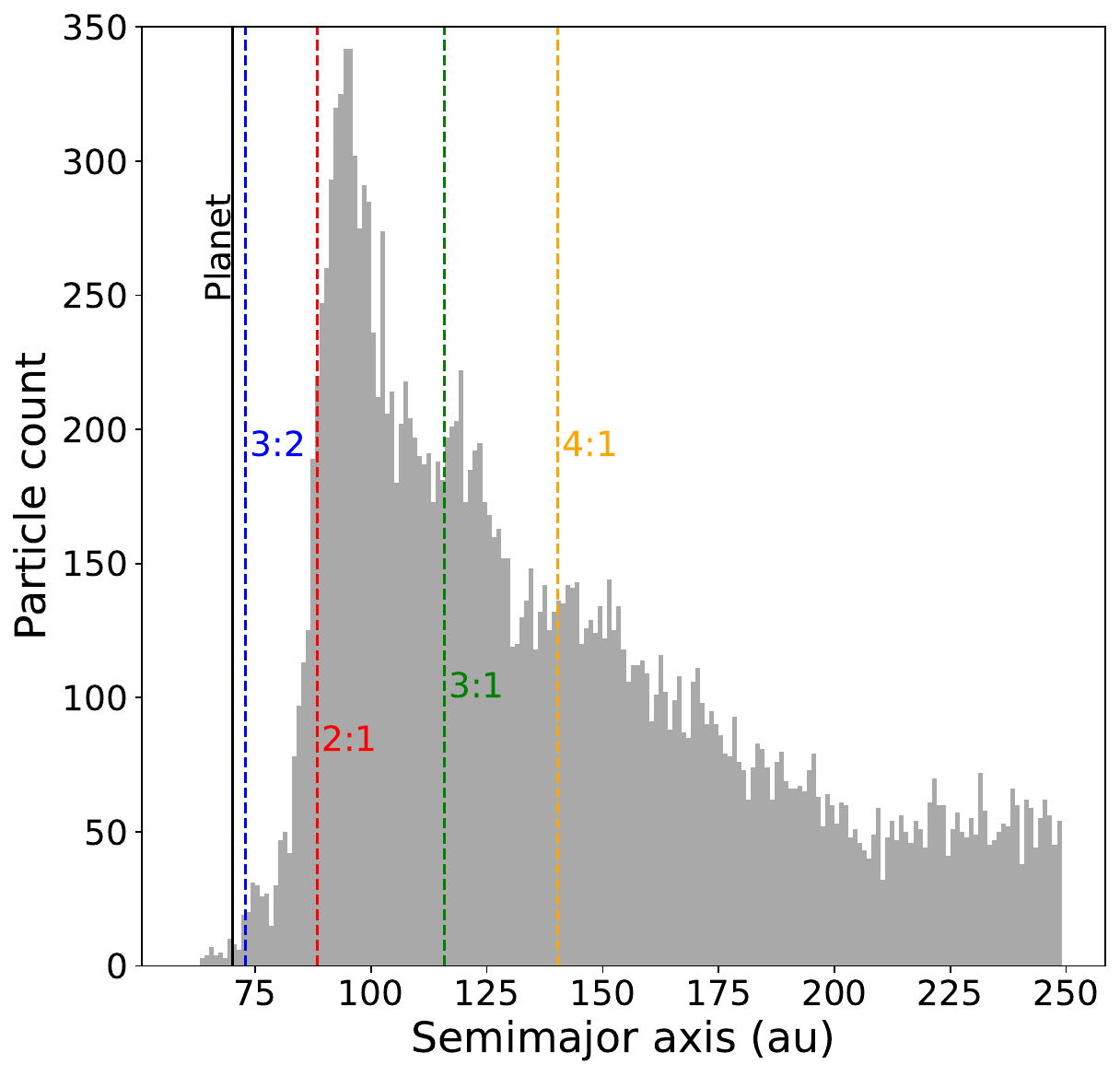}
 \includegraphics[scale=0.21]{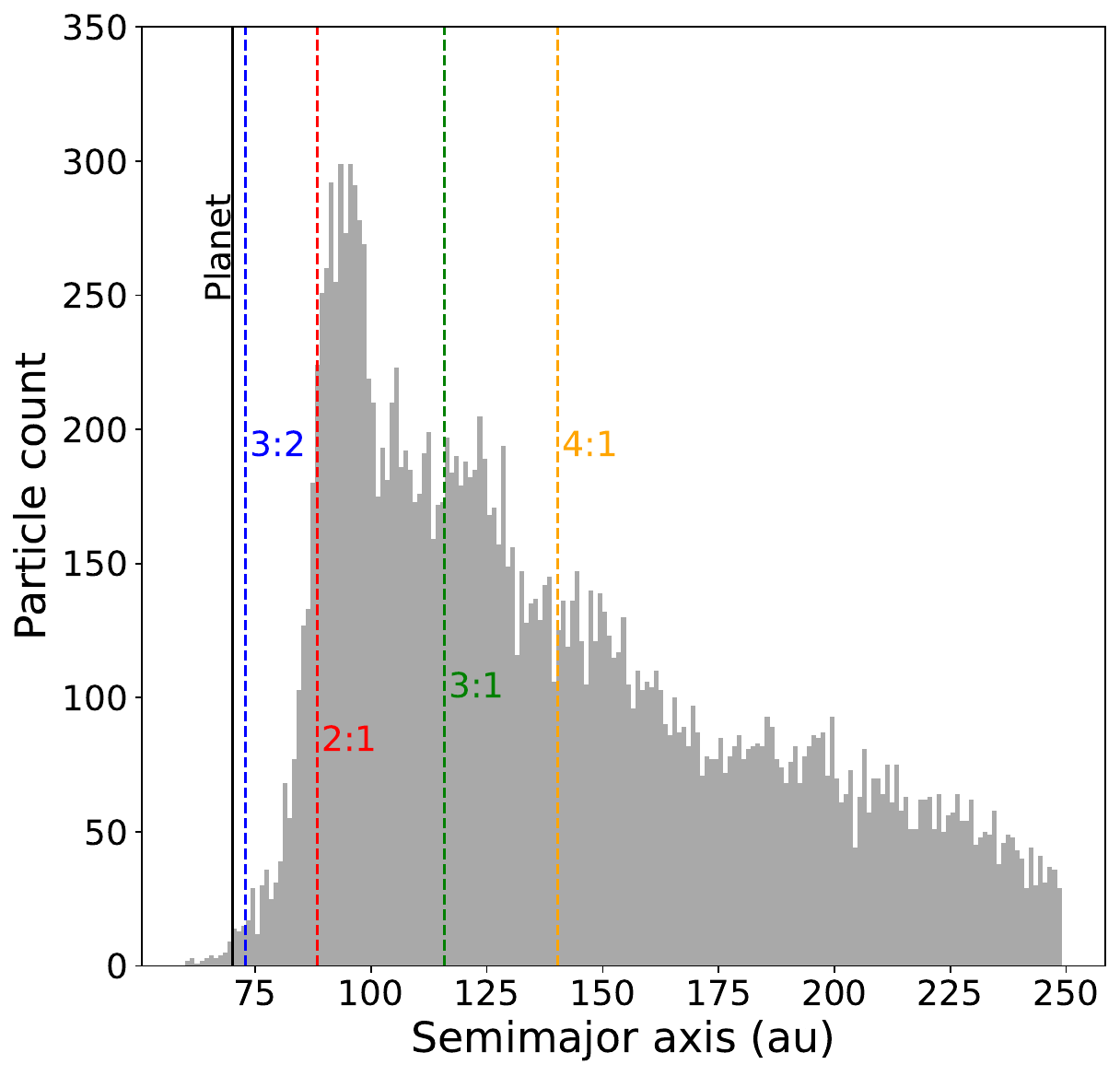}
 \caption{\label{fig:semiecchist}Evolution of newly created particles' semimajor axis at 0 years, 2~kyr, 4~kyr, and 6~kyr from left to right.  The black line indicates the position of the planet, and the blue, red,  green, and yellow vertical lines indicate the locations of the 3:2, 2:1, 3:1, and 4:1 resonances (for $\beta$=0.5) respectively.}
\end{figure*}

The evolution of the particles' semimajor axis may provide insight into the mechanics at play.  Figure~\ref{fig:semiecchist} shows the number of particles at four times during the simulation.  Here we show the 6~kyr result, but the outputs for the other timescales show similar patterns.  We examine only the first generation of particles, not a stack of continually generated particles.

The semimajor axis plots includes 4 vertical lines.  The black indicates the semimajor axis of the best-fit planet.  The blue, red, green, and yellow lines indicate the locations of the 3:2, 2:1, 3:1, and 4:1 resonances of a $\beta$=0.5 particle with the planet.  At the beginning of the simulation, there is a peak of particles at the 2:1 resonance, but this quickly migrates to show a dearth and peak just before and after respectively of the 2:1 line.  A second smaller peak appears just above the 3:1 line, and another potential small bump may exist near the 4:1.  This suggests that orbital resonances could play a role in the structure and stability of the system, as they do with planetary interactions \citep{FabryckyTremaine2007, steffenhwang2015}.

\subsection{Scale Height} \label{sec:scale_height}
The scale height $H$ of HD~181327 has been reported as 4-8~au \citep{schneider2006, stark2014, milli2024} within the parent ring.  We measure  the height of the disk in our model using the standard deviation of the vertical positions within the parent ring (74~au to 94~au).  For the 6~kyr result, we find a height of 11.1~au, somewhat larger than observed values. Our planet-based models tend to produce values at or above of previous estimates, though our most refined models (Section~\ref{sec:combanalysis}) do produce scale heights within the range of previous observations 

\begin{figure}
 \includegraphics[scale=0.35]{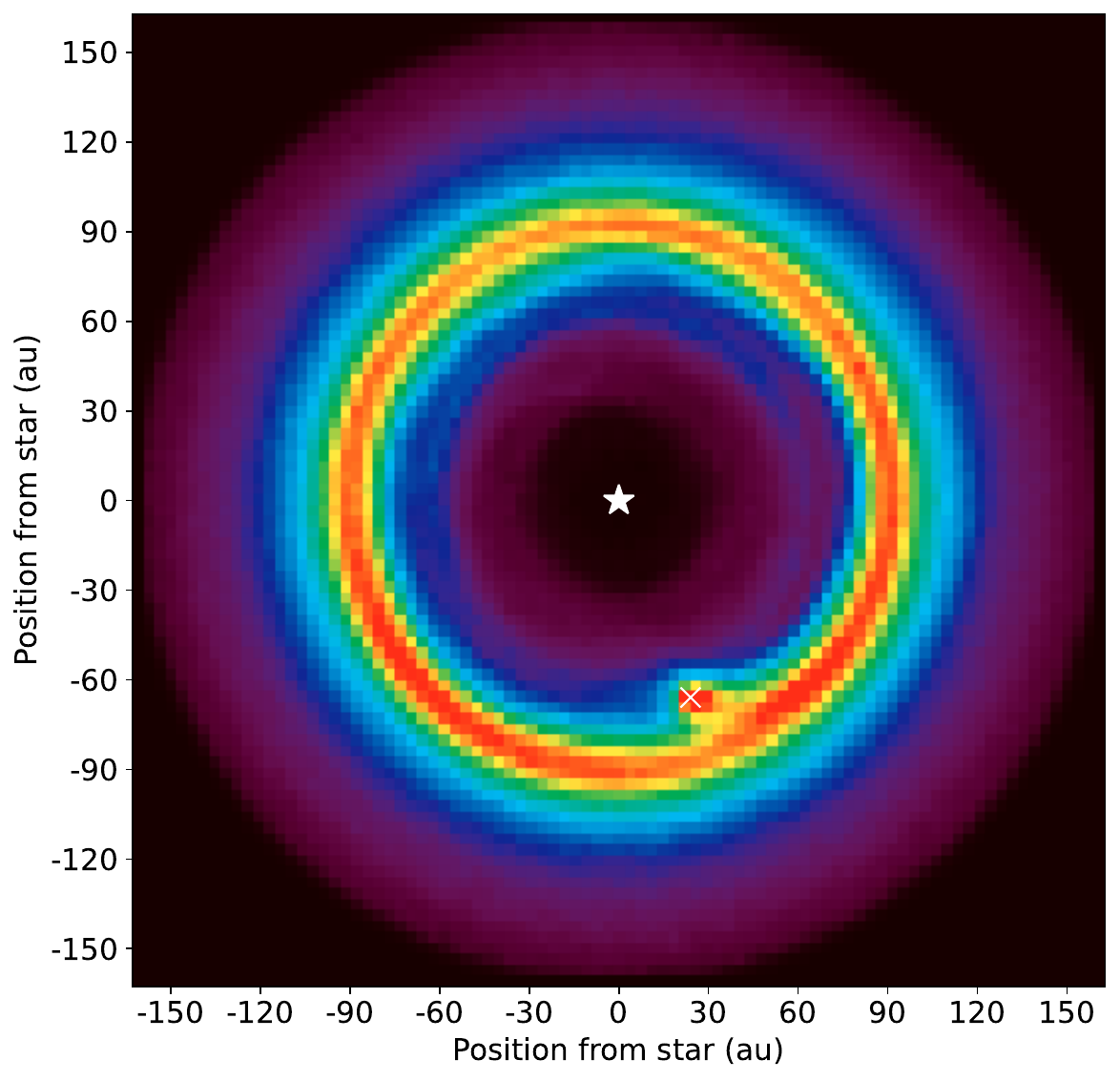}
 \includegraphics[scale=0.51, trim=0 -30 -20 0]{stark_scale_vertical_narrow.pdf}
 \caption{\label{fig:run74g4Large}Disk of large ($\beta$$\le$0.1) grains under the influence of the best-fit planet on a 6~kyr timescale (see Table \ref{tab:bestResults}).  The planetary configuration that drives small ($\beta$$\sim$0.5) particles into a high-density arc does not drive large particles into an arc.  Panels and symbols are identical to those described in Figure~2.}
\end{figure}

\subsection{Behaviour of Larger Particles}
\label{sec:largeparts}
Our results thus far have illustrated the behaviour of small ($\beta$$\approx$~0.5) particles (particle radii of $1-2$ $\micron$ for mean grain density of 2$\rm~g/cm^{3}$).  Such small particles are observable in the visible bandpass of STIS, but would less prominent or invisible at longer wavelengths.  The fact that observations at different wavelengths are sensitive to different particle sizes may provide an explanation of why observations of HD~181327 in the infrared by \citet{milli2024} and at millimeter wavelengths by \citet{marino2016} both indicated an axisymmetric disk without a pronounced arc such as reported by \citet{stark2014}.  

We next briefly examine how larger particles behave under the same conditions.  We took the best-fit planetary parameters from our 6~kyr and 20~kyr collision lifetime results (Table~\ref{tab:bestResults}) and applied them to an identically structured initial disk, but composed of larger particles with $\beta$~=~0.0001 to 0.1 (corresponding to grain sizes of 7~mm to 7~$\micron$ respectively).  

These larger particles require some changes to the initial simulation parameters.  They are less affected by radiation pressure and as a consequence, instead of being put on a parabolic orbit, a $\beta$=0.1 released with identical orbital speed as the planet will be immediately put on an $e$$\approx$0.1 orbit.  As before, a random ejection "kick" is added.  However, a $\beta$=0.1 particle is on the order of $100\times$ more massive than a $\beta$=0.5 particle.  Given our "typical" ejection kick for the best-fit run of a $\beta$=0.5 particle, what is a reasonable value for the ejection kick of a more massive particle under identical circumstances?

The ejection velocity scaling of asteroid collisions is a non-trivial topic and beyond the purview of this paper.  Such values can be stated in terms of total mass ejected given a particular velocity \citep{hyodogenda_2020,houshols2011}.  For our purposes, we need a sense of how the speeds of different masses of individual particles compare for a given collision.  \citet{nakafuji1991} performed low-speed (up to 100~m/s) experiments and found $v\propto~m^{-1/6}$, but our collision speeds are an order of magnitude higher.  If that relationship held, then we could expect a $\beta$=0.1 particle to have roughly 45\% the ejection speed of a $\beta$=0.5 for an equivalent collision event.

To account for many possibilities and give the best chance of success for replicating the Target Map with large particles, we tried several different ejection kicks, and applied them to our best-fit 6~kyr and 20~kyr simulations.  The ejection kicks we tried were 70~m/s, 250~m/s, 625~m/s, and 1250~m/s, which correspond to $\sim$5\%, 25\%, 50\% and 100\% respectively of the typical kick values found in the posteriors and best-fits (per Table \ref{tab:bestResults}).

Regardless of ejection speed (even implausible ones) and for all particle lifetime scenarios, when our previous best-fit results are applied to large particles, a matching arc of the Target Disk is not recreated.  There are significant deviations in the inner edge and extended cloud as well.  In the best-case of these scenarios (50\% kick for 6~kyr), the average $\chi^{2}$ per pixel values for entire disk, parent ring, and the arc region are 0.59, 0.50, and 0.49 respectively, all at least 4$\times$ above the values in Table~\ref{tab:bestResults}.  The result is shown in Figure~\ref{fig:run74g4Large}.  The disk is nearly axisymmetric, and is a much better match to the ALMA image from \citet{marino2016} (no asymmetry) than the Target Disk.  For HD~181327, the planet that is capable of creating a 90$^\circ$ high-density arc using $\beta$$\approx$0.5 particles does not create the same pattern with particles of $\beta$$\le$0.1.  This supports the premise that the arc observed by \citet{stark2014} is the result of $\beta$$\approx$0.5 particles, observable to STIS in the visible spectrum but not observable in longer wavelengths.

\begin{figure}
 \includegraphics[scale=0.35]{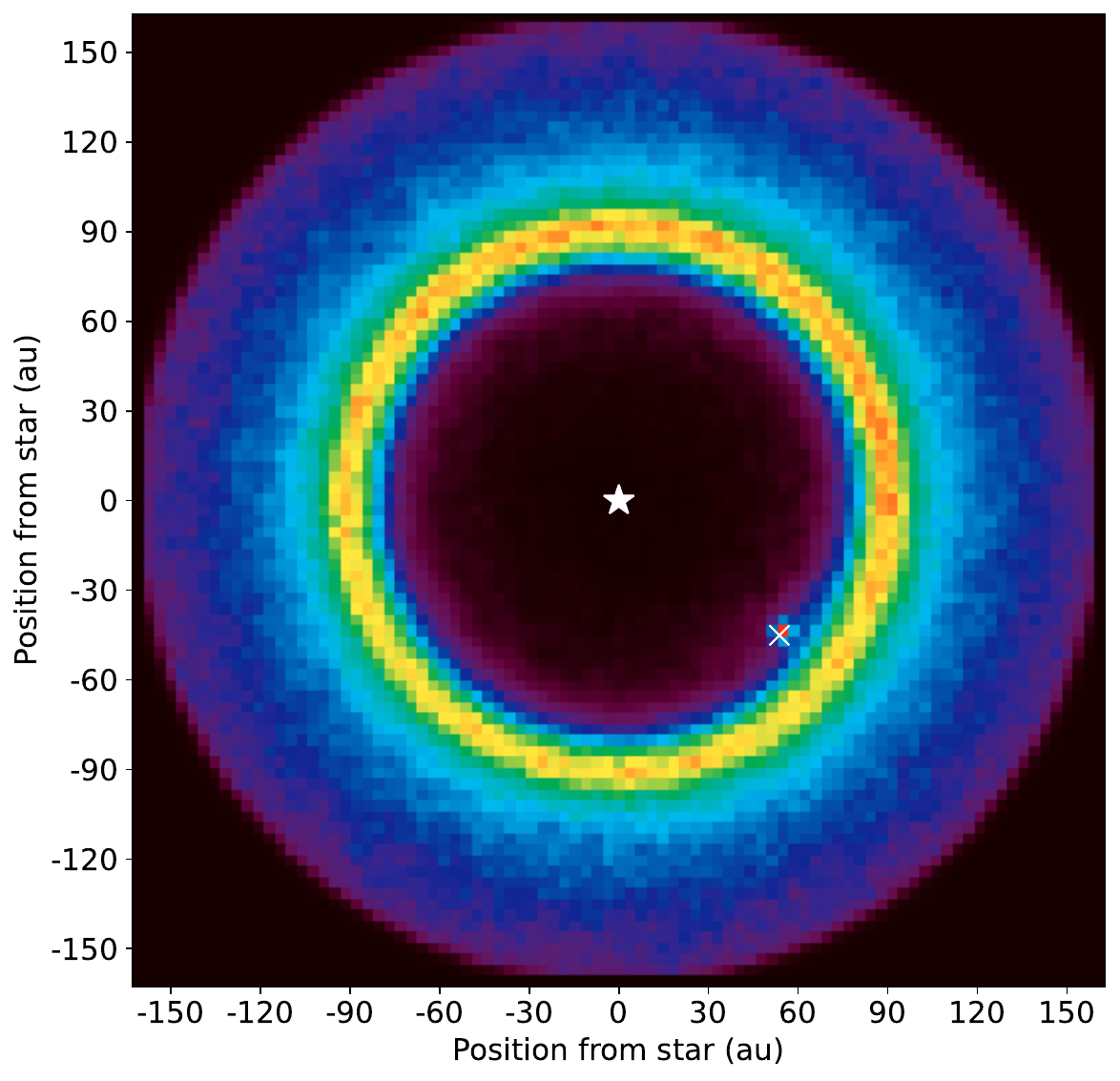}
 \includegraphics[scale=0.51, trim=0 -30 -20 0]{stark_scale_vertical_narrow.pdf}
 \includegraphics[scale=0.35]{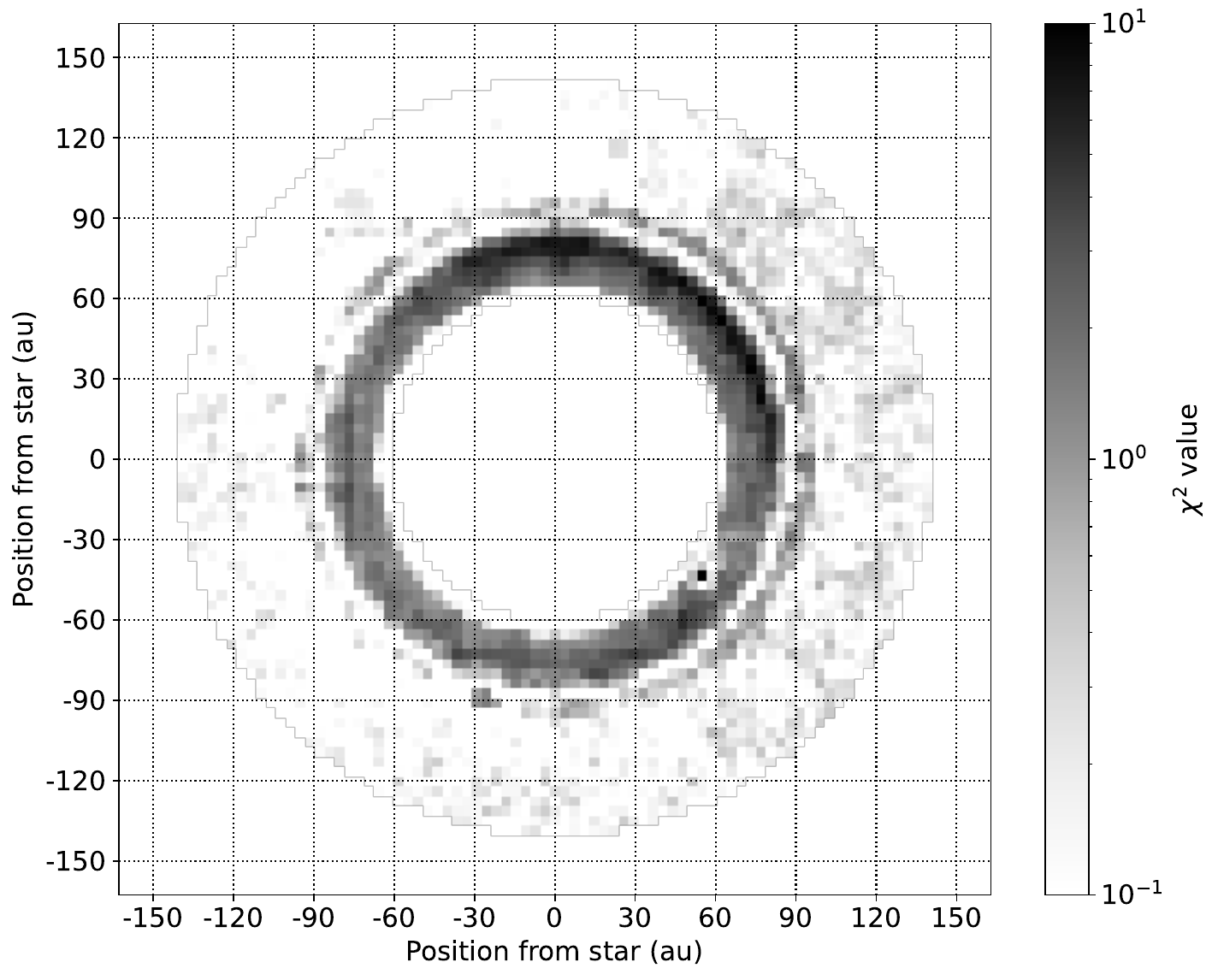}
 \caption{\label{fig:run74gAll}Best Fit result for a particle lifespan of 6~kyr, but using particles of all sizes ($\beta$=0.05 to 0.8).  The top image is the simulated optical depth map, with the white $\times$ marking the location of the planet.  At the bottom is the $\chi^{2}$ map when compared to the Target Map.  The refined disk has a mismatch on the inner edge, but otherwise produces a good match to the Target Disk.}
\end{figure}

\subsection{Further Tests and a Refined Disk Model}
\subsubsection{Combining Particles of Different Sizes / Ejection Speeds}
\label{sec:refinedmodel}
Thus far, we have examined simulations of particles with a narrow range of particle sizes, focusing on disks comprised of particles with $\beta$ values from 0.4 to 0.6.  These particles were given a specific "kick" of a particular magnitude but in a random direction.  Here we seek to take our previous results and apply them to a more realistic disk.

Our initial models were based on the hypothesis that particles with $0.4$<$\beta$<$0.6$ dominate the reflected light. To verify this hypothesis, we simulated additional disks consisting of a full range of particle sizes.  Our new disk was based on the best-fit parameters from our 6~kyr run, but using particles of $\beta$ 0.05 to 0.8.  We refer to this improved disk model as our "refined" disk.

The particle size distribution of HD~181327 has a Dohnanyi-like distribution with a size index near -3.5 \citep{lebreton2012, marino2016, milli2024}.  
Our reconstructions of the Target Disk assume equal albedo for all particle sizes and an amount of reflected light proportional to their radius squared.  We thereby account for the particles total relative scattering ability.

The ejection speed values for the new disk are also adjusted for different particle sizes, as it was in Section~\ref{sec:largeparts}.  Particles of different $\beta$ values had their kick values scaled per the $v\propto~m^{-1/6}$ law used previously. Further, instead of using a static speed for each particle size segment, every particle was given its own unique kick value, informed from our previous results.  We experimented with various distributions for these kicks.  The distribution that gave us the best result has a mean speed of 3/4$\times$ the best result (see Table~\ref{tab:bestResults}) uniformly distributed over $\pm$50\% of this value.  For example, in this new simulation, the ejection speeds of each $\beta$$\sim$0.5 particle (previous best-fit kick of $\sim$1.2~km/s) is within the range of 0.9$\pm$0.45~km/s. The speeds for each $\beta$ value are scaled accordingly.

\begin{figure*}
 \includegraphics[scale=0.3]{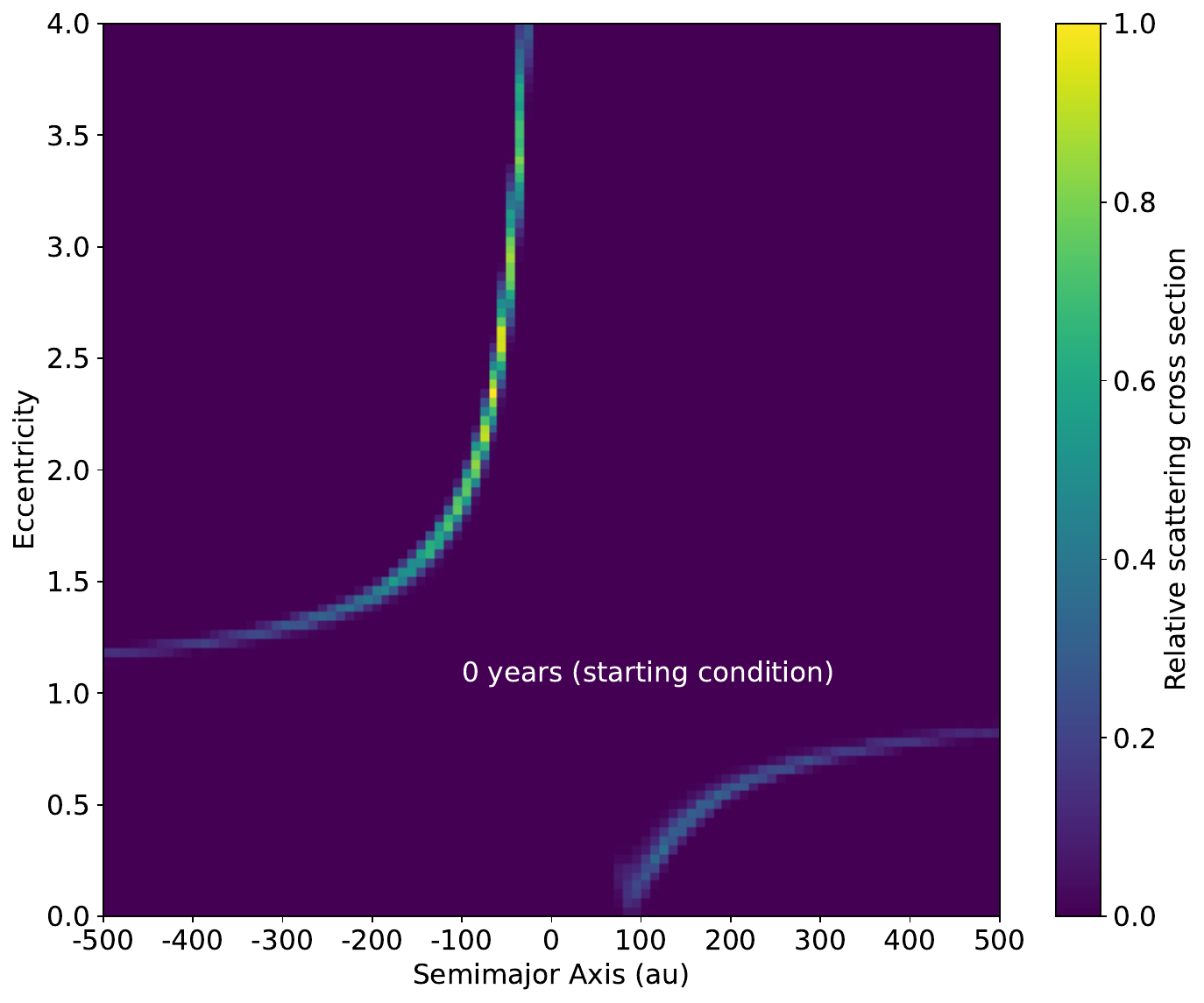}
 \includegraphics[scale=0.3]{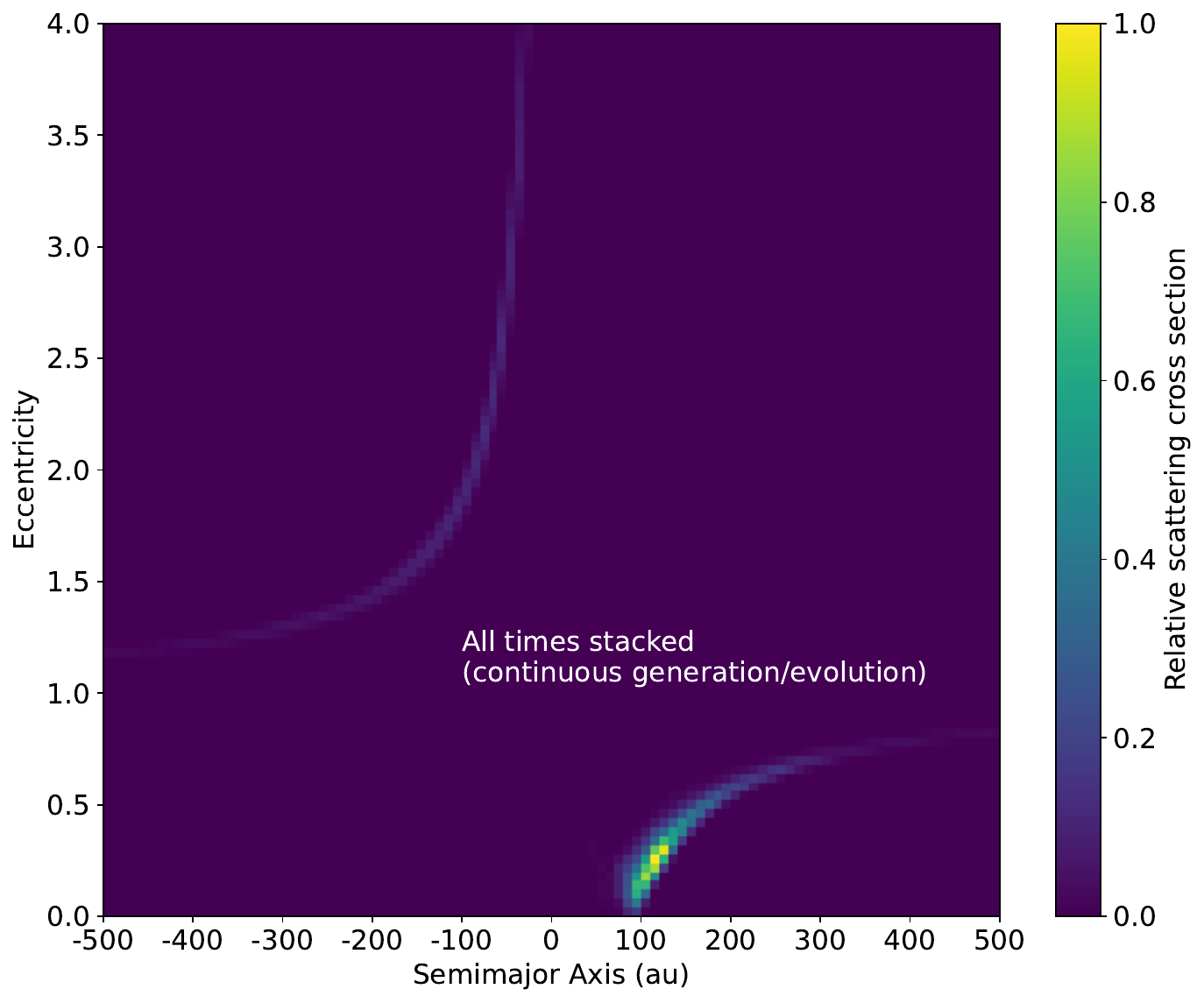}
 \caption{\label{fig:semieccevol}Heat map of semimajor axis vs eccentricity, at the beginning of the simulation and after 6~kyr.  The left panel shows the distribution of newly created particles (with ejection kick included). These are are primarily on hyperbolic orbits and quickly move outside of 150~au.  The right panel shows the stacked result, representing the ongoing distribution of particles of various ages.  Particles that are not initially lost build up in numbers and become the dominant source of scattered light, concentrated with eccentricities of 0.2 to 0.3. }
\end{figure*}

\subsubsection{Assessment of the Refined Model}
Our refined model includes a variety of particles of different sizes and ejection speeds, and is shown in Figure~\ref{fig:run74gAll}.  This scenario creates an arc of increased density with an overall strength (7.4\%) equal to our earlier model (per Table~\ref{tab:relparticles}).  However, there some differences to our previous models.  While the parent and extended disks remain a good match to the Target Disk, the $\chi^{2}$ map indicates a lower-quality fit around the inner edges of the parent disk when compared to the $\beta$$\approx$0.5 simulations (see Figure~\ref{fig:run74g}).  We also find the resulting parent ring is narrower than the Target Disk. From these results, we conclude that our model is relatively insensitive to the precise ejected velocity distribution adopted.

\subsubsection{Evolution of the Refined Disk Grains}
\label{sec:combanalysis}
The dust size distribution created by ongoing collisions within the parent disk is expected to create more small particles than large ones; however in our model the light reflected from our model disk is dominated by the particles that remain in orbit. To demonstrate this, we examined the time evolution of the grain distribution.  Figure~\ref{fig:semieccevol} shows the proportion of light scattered from the grains at the start of the simulation compared to the steady-state disk.  Newly-created grains on hyperbolic orbits initially dominate the scattering, but these grains move outside of 150~au within a few hundred years.  The few particles that are not lost accrue and become the dominant source of light.

\begin{figure*}
 \includegraphics[scale=0.25]{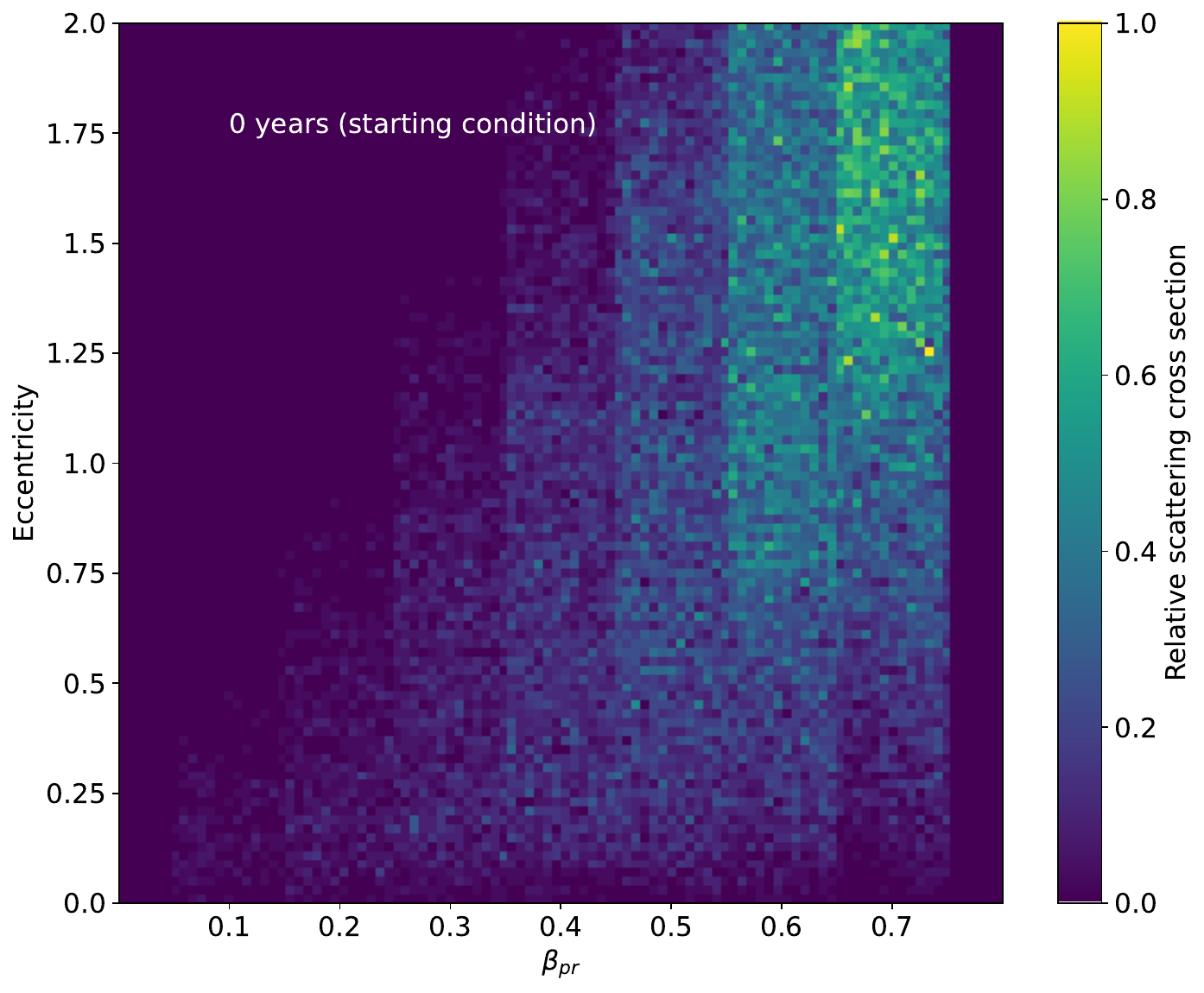}
 \includegraphics[scale=0.25]{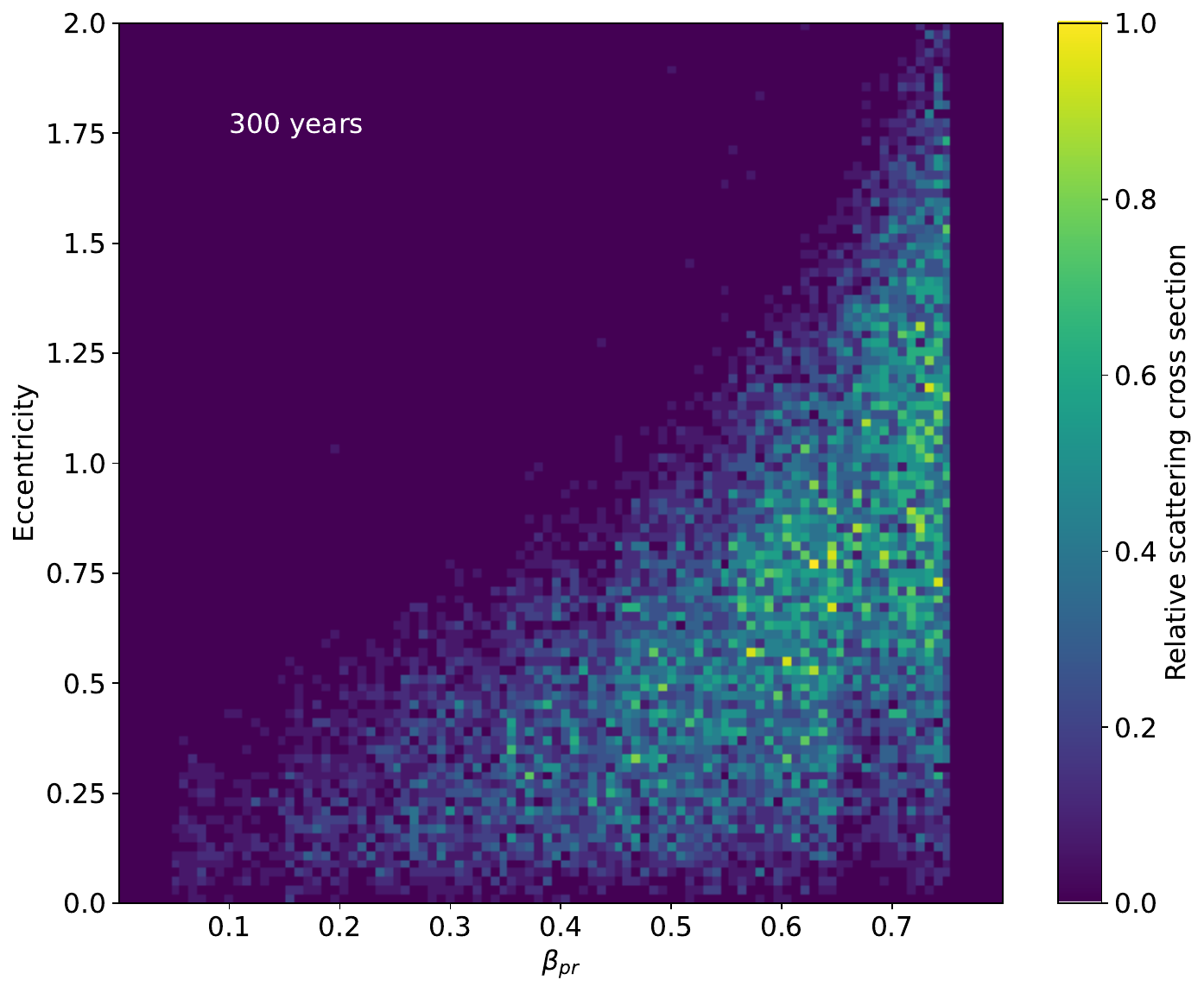}
 \includegraphics[scale=0.25]{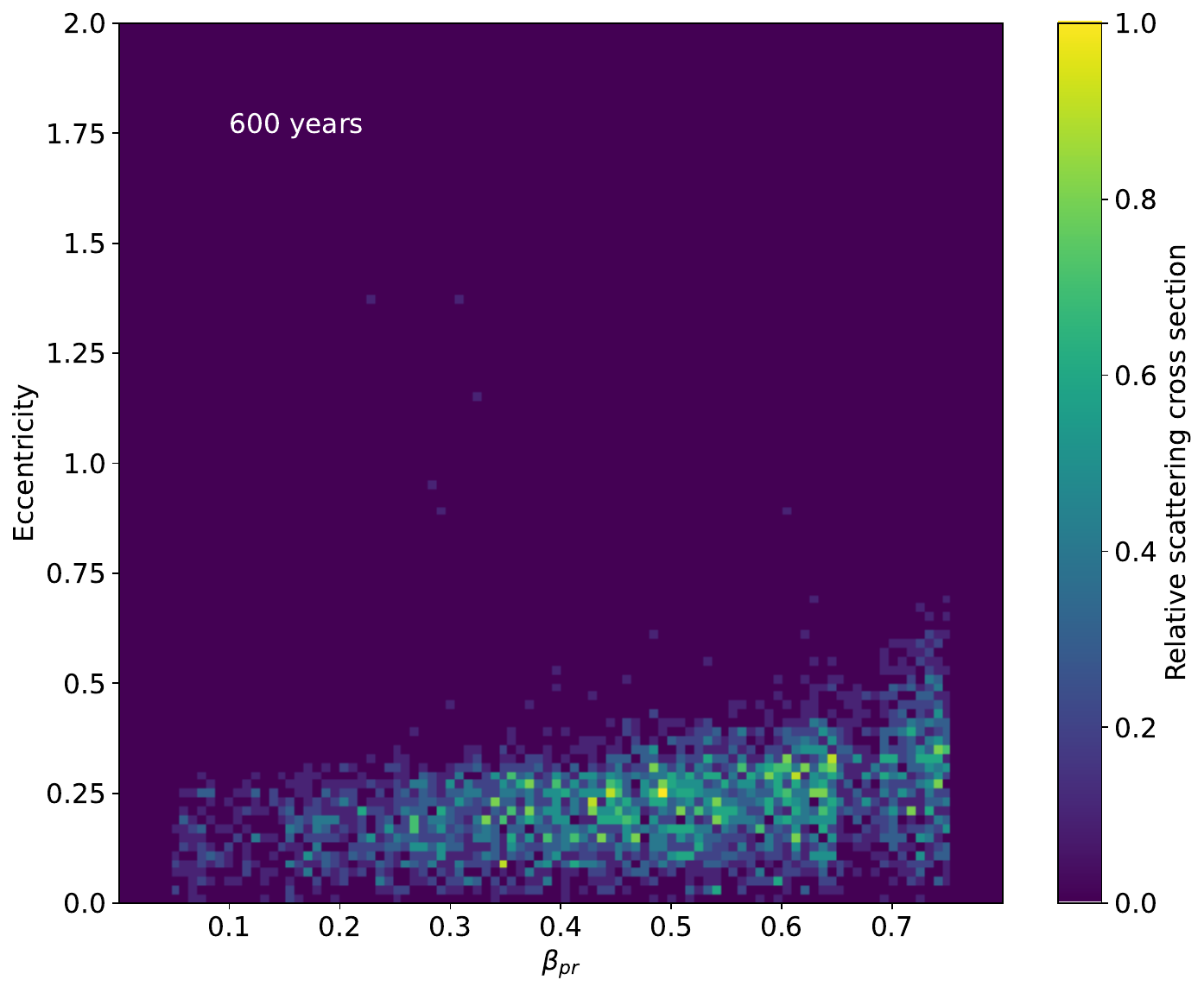}\\
 \includegraphics[scale=0.25]{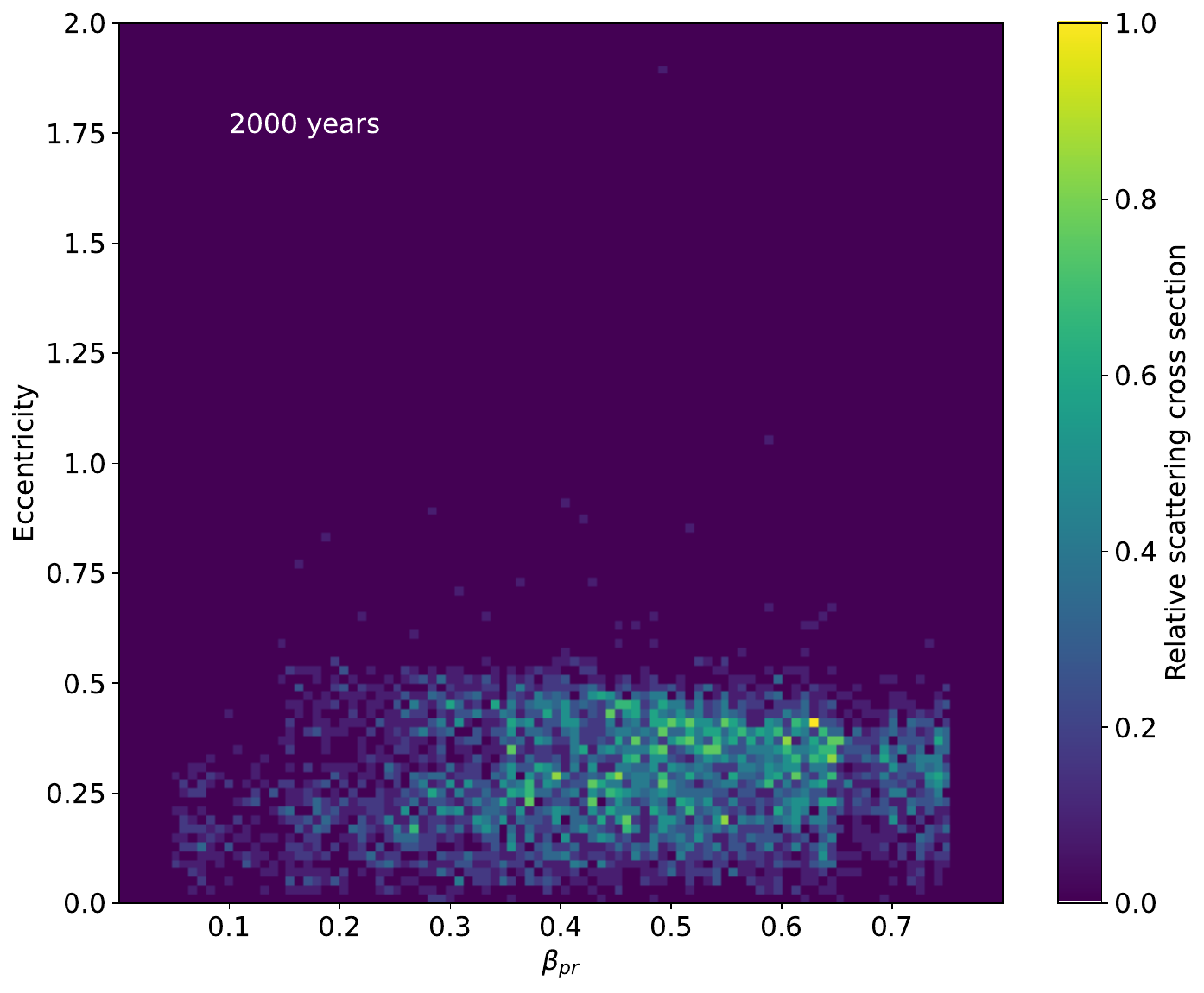}
 \includegraphics[scale=0.25]{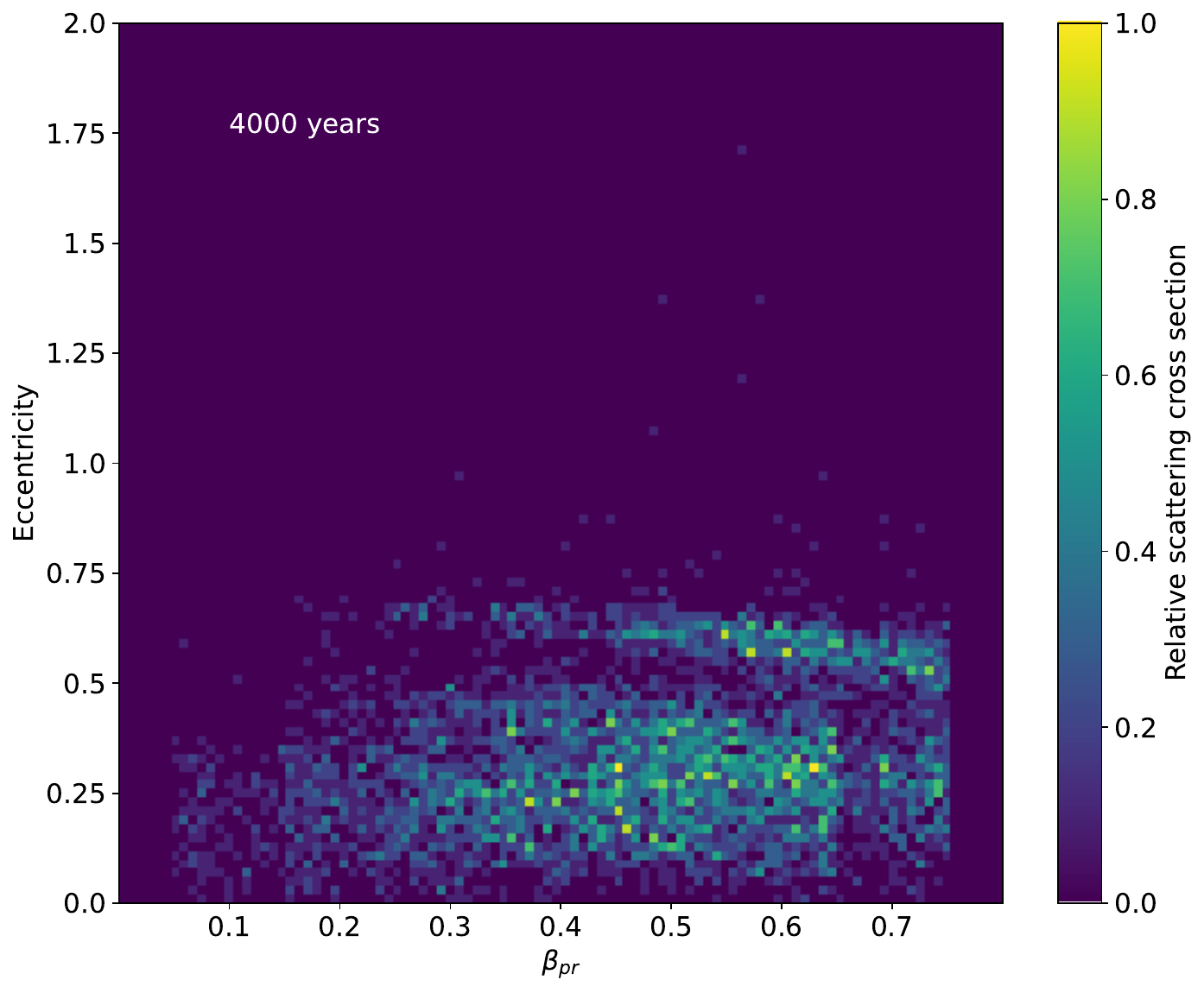}
 \includegraphics[scale=0.25]{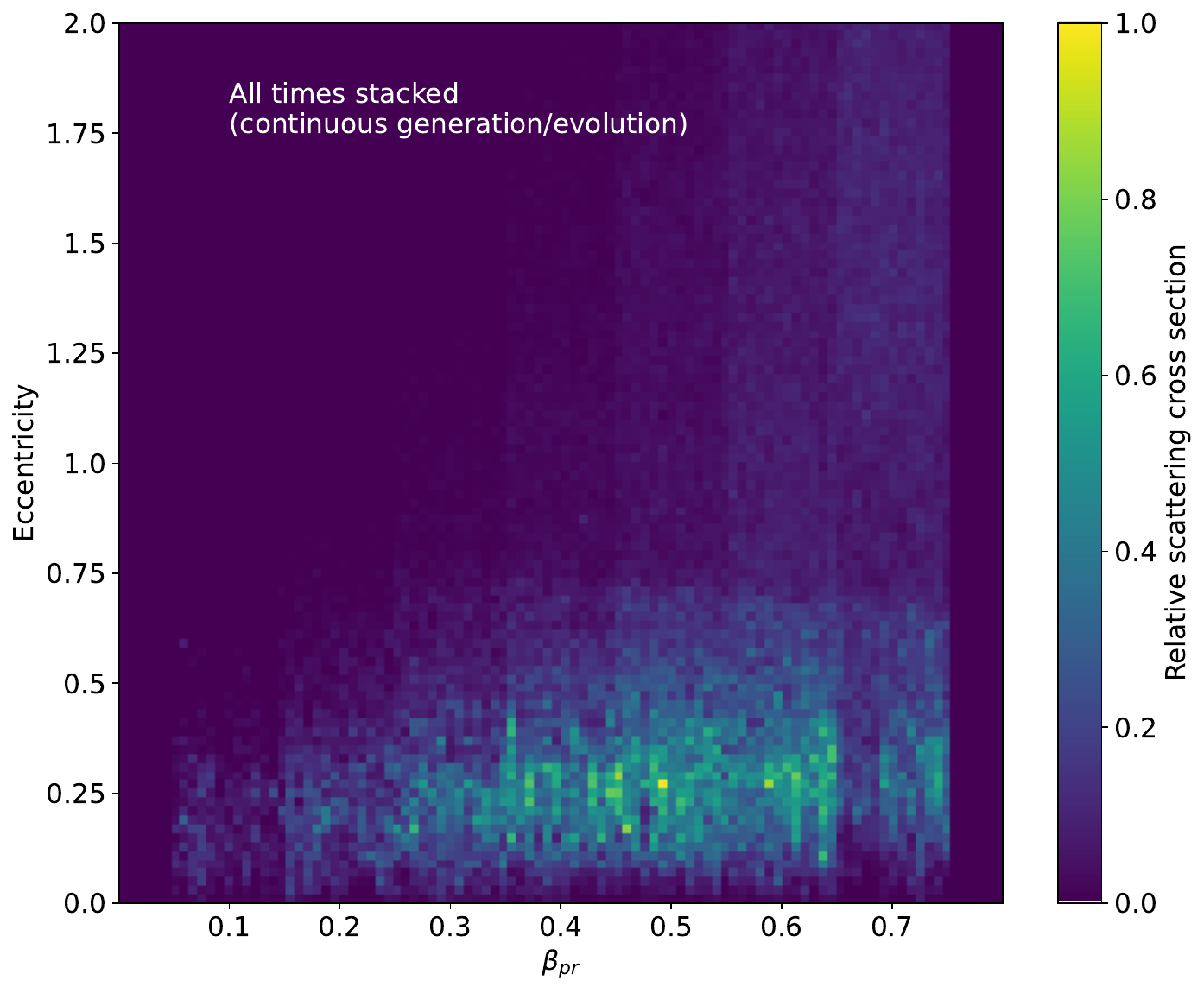}
 \caption{\label{fig:betavsecc}Heat map of the time evolution of a single generation of particles.  The maps display the scattering cross section within 150~au of the star. On the top row are snapshot of the distributions at ages of 0, 300 and 600 years respectively.  The high-$\beta$ high-$e$ particles that initially contribute the most scattering cross-section are lost quickly.  On the bottom row, the particle ages are 2000 and 4000 years.  The final panel (bottom-right) shows the stacked (all times representing the steady state generation and evolution of particles), and demonstrates the most dominant particles to the scattered light are $\beta$=0.4-0.6 with $e$=0.2-0.4.  Note that the vertical striations are artifacts of how we combined particles of different $\beta$ values in our simulations.} 
\end{figure*}

To further analyze the behaviour of the grains, Figure~\ref{fig:betavsecc} shows the time evolution of a single generation of particles for different values of $\beta$.  The smallest particles are initially the most numerous, but they are lost quickly to radiation pressure.  The mid-sized particles ($\beta$=0.4-0.6) are partially lost, but those retained on bound orbits become the largest overall contributor to the scattered light.  The last plot of this Figure~is the stacked result, representing the steady-state of particles across all ages up to their collisional lifetime.  In the steady-state, $\beta$=0.4-0.6 particles with eccentricities of 0.2-0.3 dominate the scattering.

\begin{figure} 
 \includegraphics[scale=0.4]{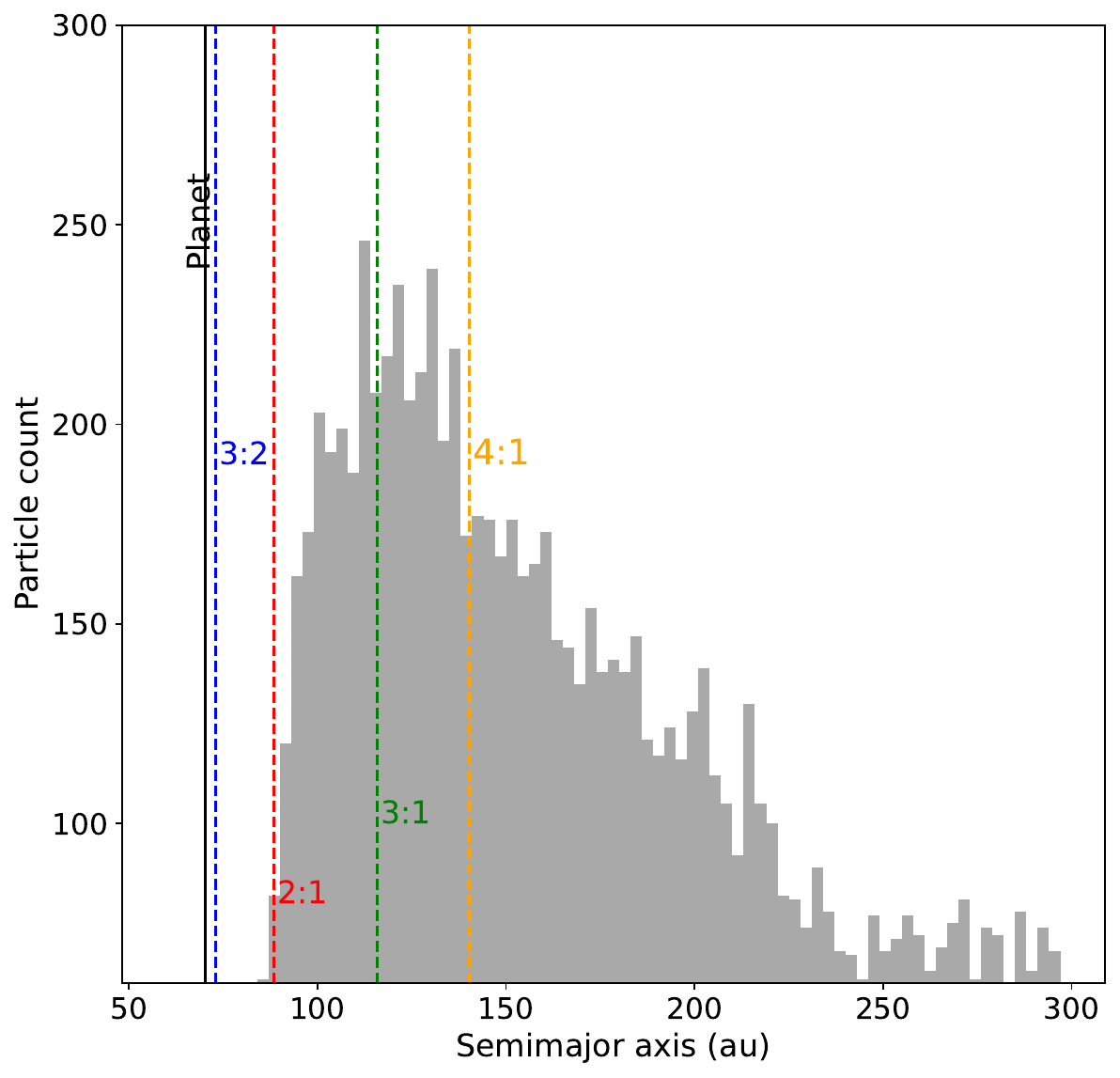}
 \caption{\label{fig:semi6kyrref}Semimajor axis distribution of the particles in the refined disk after 6~kyr.  The black line indicates the position of the planet, and the blue, red,  green, and yellow vertical lines indicate the locations of the 3:2, 2:1, 3:1, and 4:1 resonances (for $\beta$=0.5) respectively.}
\end{figure}

The specific particles in the arc are always changing, and their resonance properties are difficult to determine, but similar arcs have also been seen by \citet{liou1999} in models of Kuiper belt dust trapping by the giant planets. \citet{liou1999} models use relatively gentle ejection speeds, placing dust on $e$=0.1 and $i$=$10^\circ$ orbits, which may account for the differences seen.  Arcs like those found here (asymmetrical with respect to the planet) have been seen in \citet{dermott1994} in models of the Earth's dust ring, related to trapping of in-spiraling asteroid dust near 1:1 resonances; however the grains in our ring are well outside the 1:1 resonant zone. 

The particles in the region of the arc have perihelia concentrated at 81-84~au, and median eccentricities ranging from 0.14 for $\beta$$\approx$0.1 particles and up to 0.45 for $\beta$$\approx$0.7.  Figure~\ref{fig:semi6kyrref} shows the semimajor axis distribution of the particles in the refined disk at 6~kyr, which is broader for the refined disk compared to our earlier results.  Instead of the distribution peaking near the 2:1 period ratio (see Figure~\ref{fig:semiecchist}), the distribution peaks between the 3:1 and 4:1 resonances.  The 2:1 shows less than half the number of particles as the 3:1, and there are very few near the 3:2.  This trend applies for all bound particle types, not just the $\beta$=0.4-0.6 grains.  This suggests the revised velocity-kick has distributed the particles more broadly, resulting in more of the inner edge particles being spread out across the disk.  While the particle distribution changes as it ages and has various over- and under-densities throughout its circumference during its lifetime, starting around 3~kyr the primary arc (western quadrant) becomes apparent and holds its position with respect to the planet.  The individual grains in the arc themselves are not stationary with respect to the planet, and move from perihelia within the parent ring to the more diffuse outer disk and back. 

\subsubsection{Scale Height of the Refined Disk} \label{sec:ref_scale_height}
The scale height of our refined disk parent ring (74~au to 94~au) is found to be 7.6~au, as estimated from the standard deviation of the vertical positions of particles within the parent disk.  Thus, unlike our initial model value (11.1~au), our refined model has a scale height consistent with the analysis from previous studies \citep{schneider2006, stark2014, milli2024}.  The reason for the difference between our refined disk and the previous is likely the lower "kick" speed and distribution.  Our refined model used kick speeds of 0.45-1.35~km/s, while the original used a static 1.2~km/s.

\section{Other Considerations}
\subsection{Model Limitations}
\label{sec:modellimits}
Our simulated model has several limitations.  First, it presumes a single type of particle.  The composition of the debris disk of HD~181327 includes many dust species (carbon-, iron-, silicon-based) along with water ice \citet{lebreton2012, Xie2025}.  We have also assumed that particles of all sizes have the same albedo.  However, depending on the distribution and albedo of these various species, different grain sizes may contribute to the brightness profile differently.

Second, our simulations use ejection "kicks" to simulate the ejected debris from a collision.  The magnitude of these kicks is $\sim$1~km/s, corresponding to the relative velocity of two particles of $e$$\approx$0.3.  This is in mild tension with the usual assumption that $e$$\sim$i$\sim$$H$, since the scale height $H$ of HD~181327 has been measured to be $\las$$0.1$.  However, our speed distribution 1) has a scale height $H$$\approx$$0.1$ (Section~\ref{sec:ref_scale_height}, which is of the same order as the measured values from \citet{schneider2006, stark2014}, 2) contains a distribution of eccentricities peaking between 0.2-0.3 (see Fig~\ref{fig:betavsecc}), and 3) the appearance of the observed disk is well-reproduced by our model.

We note that while collision speeds of this magnitude are plausible within our model, the resultant speed of ejected debris is more difficult to determine.  Our refined model from Section~\ref{sec:refinedmodel} requires a mean ejection speed of less than 1~km/s, which reduces this tension.  Particles with $\beta$>0.5 and with low ejection speed are naturally lost (immediately put on $e$$\approx$1 or $e$>1 trajectories).  Only grains with sufficient ejection speeds in the proper direction remain bound and thus they contribute disproportionately.  This accumulation means that each generation of grains needs only a small fraction with high ejection speeds.  The ejection speed of debris is a complex topic beyond the scope of this paper, but if ejecta speeds of this magnitude can exist (by whatever mechanism) in modest quantities, then a planetary explanation for the arc observed by \citet{stark2014} is viable.

\subsection{Detection Limits of Planet Masses}
\label{sec:detectlimit}
There have been previous estimates and limits placed on the mass and orbital distance of a potential planet within the HD~181327 system.  As part of the Gemini NICI planet-finding campaign (using direct imaging), \citet{wahhaj2013} estimated the that a 4.0~$\rm M_{J}$ planet in HD~181327 would be detectable at 1.0" from the central star and a 2.6~$\rm M_{J}$ planet at 2.0".  Interpolating from their limits, the detectability our best-fit planetary distance (70~au, or 1.4") would be $\sim$3.5~$\rm M_{J}$.  This suggests that our predicted planet must have a mass lower than 3.5~$\rm M_{J}$, otherwise the Gemini NICI campaign would have observed it.  This informed the results we used throughout Section~\ref{sec:results}.  Our MultiNest results suggest that more massive planets could induce the arc, but because of the Gemini NICI results, we focused on solutions that were near or below the limits set by \citet{wahhaj2013}.

In another study, \citet{rodigas2014} through dynamical N-body simulations made predictions for shepherding planets in the interior of several debris disks.  For HD~181327, they found the planet could have a mass no larger than 15~$\rm M_{J}$ and must be at least 35~au from the star, and our results are consistent with these limits.

Our simulations point to a planet with a semimajor axis of $\approx$70~au and a range of possible masses, 2.0 to 4.0~$\rm M_{J}$.  Our posteriors are inside the limits set by \citet{rodigas2014} and \citet{wahhaj2013}, though our upper uncertainties exceed these limits.  Most of our best-fit masses are near the Wahhaj limit, but a few have planetary masses as low as 1.6~$\rm M_{J}$ that are also capable of producing a similar arc.  Thus our results adhere to the pre-established detection limits.

\subsection{Longer Timescales and Particle Lifetime Scaling}
Our refined disk model has focused on using the best-fit result from the 6~kyr result from Section~\ref{sec:bestfits} (Figure~\ref{fig:run74g}).  As an experiment, we permitted this run to run out to 20~kyr.  This resulted in the loss of the arc, with a very axisymmetric parent ring.

We note that particles of different sizes have different lifetimes.  Work by \citet{ThebaultAugereau2007} indicate that there is a sharp minimum lifetime at approximately 10$\times$ the blowout radius ($\beta$$\sim$0.05).  Per Equation~7 from \citet{ThebaultAugereau2007} a $\beta$=0.1 particle has a lifetime $\sim$25$\times$ less than a $\beta$=0.5 particle.  Further, grains contribute less to the scattering the larger they become ($\propto$$s^{-1.5}$).  Combining these two effects, we expect that regardless of the exact lifetime, large particles will have a small impact on the disk profile.

\subsection{Rotation Direction}
The motion of orbits in our simulations shown in Section~\ref{sec:results} is in the clockwise direction, consistent with analysis by \citet{marino2016}.  The disk and arc are very symmetric across the positive diagonal, so we expected the direction of rotation to play a minimal roll in our results.  For confirmation, we also ran simulations in the counter-clockwise direction.  We find that we get statistically identical results (near-identical posteriors).  If the disk actually rotates in the counter-clockwise direction our results would still stand, but the simulated maps in Section~\ref{sec:bestfits} and location of the simulated planet (the $\times$ symbol) would be flipped across the positive diagonal.

\section{Summary}
The primary goal of this study was to determine whether a planet could be the cause of the asymmetry in the HD~181327 debris disk reported by \citet{stark2014}, and if so to parameterize such a planet.  We find that a planet can indeed create an arc similar to that observed.  We summarize the relevant requirements and results below.
\begin{itemize}
     \setlength{\labelwidth}{4em}
     \setlength{\labelsep}{0.2em}
     \setlength{\itemsep}{3pt}
     \setlength{\leftmargin}{1em}
     \setlength{\rightmargin}{2em}
     \setlength{\itemindent}{1em} 
  \item Grains are produced throughout the disk from an exo-Kuiper belt, with approximate limits of 80~au to 100~au.
  \item The likely lifetime of micron-sized grains is somewhere between 4~kyr and 8~kyr.
  \item Contributing grains are initially released on moderate-eccentricity orbits, requiring ejection speeds near 1~km/s in a retrograde direction relative to the planet, the result of energetic collision processes.  
  \item The majority of grains are released with low speeds and put on high eccentricity orbits.  These and are either lost quickly or on such large orbits that they do not often contribute to the arc.
  \item The planet has a mass of 2-4$\rm M_{J}$ and semimajor axis of 68-73~au.
  \item Only grains with $\beta$$\approx$0.5 are quickly driven to the requisite pattern.  This may explain why the arc is reported at visible wavelengths but is not at longer wavelengths.
\end{itemize}

The structure of the HD~181327 disk as observed by \citet{stark2014} is reproducible by the presence of a giant planet within the system. Our simulations are able to produce a debris disk very similar to the disk of HD~181327 with a range of planetary masses and timescales.  These simulations are premised on the existence of ongoing collisions (as indicated by \citet{Xie2025}) with adequate energy to create and eject grains of sufficient speed and moderate eccentricities in modest quantities.  While the bound grains are relatively few in number for each generation, their accrual enables them to dominate the light scattering over the unbound grains.  This hypothesis can explain the absence of structure seen at longer wavelengths, as larger particles do no take on the same distinct arc as smaller particles.  We encourage further observations of the system to corroborate or refute the planet hypothesis.

\section{Acknowledgements}
We thank the anonymous reviewer(s) for their thorough feedback that significantly improved this work.
This work was funded in part by the Natural Sciences and Engineering Research Council of Canada Discovery Grants program (Grant no. RGPIN-2024-05200).

\section{Data Availability}
This paper is based on the work of \citet{stark2014}, and all source data comes directly from that work.  No new data were generated in support of this research.



\bibliographystyle{mnras}
\bibliography{citations}





\bsp	
\label{lastpage}
\end{document}